\documentclass{jfm}

\usepackage{graphicx}
\usepackage{natbib}
\usepackage{epsfig} 
\usepackage{epstopdf}
\usepackage{amsmath}
\usepackage{subfigure}

\let\realverbatim=\verbatim
\let\realendverbatim=\endverbatim
\renewcommand\verbatim{\par\addvspace{6pt plus 2pt minus 1pt}\realverbatim}
\renewcommand\endverbatim{\realendverbatim\addvspace{6pt plus 2pt minus 1pt}}
\makeatletter
\newcommand\verbsize{\@setfontsize\verbsize{10}\@xiipt}
\renewcommand\verbatim@font{\verbsize\normalfont\ttfamily}
\makeatother

\ifCUPmtlplainloaded \else
  \checkfont{eurm10}
  \iffontfound
    \IfFileExists{upmath.sty}
      {\typeout{^^JFound AMS Euler Roman fonts on the system,
                   using the 'upmath' package.^^J}%
       \usepackage{upmath}}
      {\typeout{^^JFound AMS Euler Roman fonts on the system, but you
                   don't seem to have the}%
       \typeout{'upmath' package installed. jfm.cls can take advantage
                 of these fonts,^^Jif you use 'upmath' package.^^J}%
      }
  \else
  \fi
\fi

\ifCUPmtlplainloaded \else
  \checkfont{msam10}
  \iffontfound
    \IfFileExists{amssymb.sty}
      {\typeout{^^JFound AMS Symbol fonts on the system, using the
                'amssymb' package.^^J}%
       \usepackage{amssymb}%

      }{}
  \fi
\fi

\ifCUPmtlplainloaded \else
  \IfFileExists{amsbsy.sty}
    {\typeout{^^JFound the 'amsbsy' package on the system, using it.^^J}%
     \usepackage{amsbsy}}
    {}
\fi

\newsavebox{\astrutbox}
\sbox{\astrutbox}{\rule[-5pt]{0pt}{20pt}}

\newcommand{\pa}[2]{\fracd{\partial #1}{\partial #2}}  
\newcommand{\pdd}[2]{\fracd{\partial ^2#1}{\partial #2^2}}  
\newcommand{\fracd}[2]{\displaystyle\frac{#1}{#2}} 
\newcommand{\simd}[2]{\begin{array}[t]{c}
                                \sim \\[-.0cm]
                                \scriptstyle{#1 \rightarrow #2} 
                        \end{array}}

\def \eps{\varepsilon}
\def \om{\omega} 
\def \Om{\Omega} 

\def \trho{\tilde{\rho}}
\def \brho{\bar{\rho}}

\def \tr{\tilde{r}}
\def \hr{\hat{r}}

\def \bchi{\bar{\chi}}
\def\bpsi{\bar{\psi}}

\def\Bchi{{\mathcal X}}
\def\BF{{\mathcal F}}

\def\bbf{\mathbf{f}}

\def\bI{\mathbf{I}}
\def\bJ{\mathbf{J}}

\def \be {\begin{equation}}
\def \ee {\end{equation}}
\def \ba {\begin{array}}
\def \ea {\end{array}}
\def \bea {\begin{eqnarray}}
\def \eea {\end{eqnarray}}
\def \bse {\begin{subequations}}
\def \ese {\end{subequations}}
\def \bsea{\begin{subeqnarray}} 
\def \esea{\end{subeqnarray}}

\title[Libration-induced mean flow  in a spherical shell]{Libration-induced mean flow  in a spherical shell}

\author[A. Sauret \& S. Le Diz\`es]
{Alban\ns SAURET \and St\'ephane\ns LE  DIZ\`{E}S\footnote{Email adress for correspondance: ledizes@irphe.univ-mrs.fr}}

\affiliation{Institut de Recherche sur les Ph\'enom\`enes Hors \'Equilibre,
CNRS \&   Aix-Marseille University,\break
49, rue F. Joliot-Curie, F-13013 Marseille, France}

\date{28 November 2012}

\begin{document}

\label{firstpage}
\maketitle

\begin{abstract}
We investigate the flow in a spherical shell subject to a time harmonic oscillation of its rotation rate, also called longitudinal libration,  
when the oscillation frequency is larger than twice the mean rotation rate. In this frequency regime, 
no inertial waves are directly excited by harmonic forcing. We show however that it can generate through non-linear interactions 
in the Ekman layers a strong mean zonal flow in the interior. 
An analytical theory is developed using a perturbative approach in the limit of small libration amplitude $\eps$ and small Ekman number $E$. The mean flow is found to be at leading order an azimuthal flow which scales as the square of the libration amplitude and only depends on the cylindrical-radius coordinate. The mean flow also exhibits a discontinuity across the cylinder tangent to the inner sphere. We show that this discontinuity can be smoothed through multi-scale Stewartson layers. The mean flow is also found to possess a weak axial flow  which scales as $O(\eps^2 E^{5/42})$ in the Stewartson layers.  The analytical solution is compared to axisymmetric numerical simulations and 
a good agreement is demonstrated.   
\end{abstract}



\section{Introduction}

Most astrophysical bodies, in addition to their rotation, are subject to harmonic forcing such as precession, tidal deformation and latitudinal/longitudinal librations. The nonlinear corrections induced by such a forcing can lead to important steady and axisymmetric flow, called zonal flow, in liquid core of telluric planets, in subsurface ocean of some satellites or in atmosphere of gaseous planets. In this study, we focus on longitudinal libration which corresponds to an harmonic oscillation of the rotation rate. These oscillations are generally induced by gravitational coupling between an astrophysical body and its main gravitational partner around which it orbits \cite[][]{comstock2003}. The librational forcing can have a non-negligible contribution in the internal dynamic of a planet and a better knowledge of this dynamic can bring important information on the structure of a planet, as for example an estimation of the thickness of the fluid layer under the ice shell in some satellites like Europe, Ganymede or Encelade \cite[][]{spohn2003,lorenz2008, vanhoolst2008,rambaux2011} or  the existence of an inner liquid core in telluric planets like  Mercury \cite[][]{margot2007}.

In the present work, we  derive rigorous results for fast longitudinal libration when the libration frequency is at least twice larger than the spin frequency. Such a configuration
is not the most common situation as the main libration frequency often corresponds to the orbital frequency which is generally equal or smaller than the spin frequency. Here are
a few situations which could exhibit fast libration. For example, one could imagine a planet or a satellite for which  spin and orbital frequencies are in resonance with a 
ratio 1:2. This situation which is in principle possible \cite[][]{Goldreich1966,Wieczorek2012}  would lead to fast libration with a libration frequency equal to twice the orbital frequency. 
Fast libration could also be present in a spin-orbit synchronized object when the forcing is induced by another satellite or planet with a larger orbital frequency. 
The Galilean satellites (Io, Europa, Ganymede)  are in this situation.  Io has an orbital frequency twice larger than Europa, and four times larger than Ganymede. 
It could  a priori then force the libration of Europa and of Ganymede with a frequency twice and four times larger
 than their spin frequency respectively.  
 In exoplanet systems, a similar situation can also be found. For instance, the two first planets in the 55 Cancri system 
 have orbital frequencies of 2.81 and 14.65 days \cite[][]{Fischer2008}. Since both planets are expected to be spin-orbit synchronized,  the frequency of libration of the second planet induced by the motion of the first planet would be $\om=14.65/2.81\approx 5.2$.

The first analytical study of nonlinear effects generated by  harmonic forcing   was performed by \cite{busse1968} in the case of precession in a spheroid. He showed that a zonal flow is generated in the bulk as an effect of the nonlinear interactions in the viscous boundary layers.  \cite{wang1970} studied theoretically and experimentally the nonlinear effects associated with an oscillation of the rotation rate in a cylindrical container (see also \cite{busse2010b}). The case of a fixed tidal deformation was considered by \cite{suess1971} who argued that a counter-rotating vortex should exist near the axis of rotation.  
\cite{aldridge1969} analyzed librating spheres but their experimental study focused on the linear response in the presence of inertial waves: they showed using pressure measurements that inertial waves can be resonantly excited in a librating sphere for particular libration frequencies. This has been confirmed numerically by  \cite{rieutord1991}. In his thesis, \cite{aldridgephd} also mentioned that when the inertial waves are not significant, the flow in the bulk seems to drift with an amplitude proportional to the square of the libration amplitude but he did not report any quantitative measurement.  \cite{tilgner1999} investigated numerically the linear response to the librational forcing in a spherical shell and studied the effect of the presence of an inner core but he focused on the inertial waves and the attractors only. 
Recently,  \cite{calkins2010} studied numerically the mean zonal flow in the same geometry.  They confirmed the presence of a mean zonal flow independent of the Ekman number and which scales as the square of the libration amplitude. However, their study remained mainly limited to a fixed frequency. Some LDV measurements of the mean zonal flow have also been performed by \cite{noir2010} in a cylindrical geometry. In this geometry,  \cite{sauret2012} analyzed numerically the influence of the libration frequency. They found that the mean zonal flow was well described by the theory of \cite{wang1970} as long as no inertial waves are excited.
The only theoretical study of the zonal flow in a spherical geometry is the recent work by  \cite{busse2010}.  He obtained by asymptotic methods an analytical prediction of the mean zonal flow in the limits of 
small libration frequencies and small Ekman numbers. 
The present paper  extends this theory to the case of  a  spherical shell for any frequency.

The study implicitly assumes the absence of instability. Both shear and centrifugal instability are a priori possible when the amplitude of libration becomes sufficiently large.
Centrifugal instability has been analysed in details in a cylindrical geometry by \cite{Ern1999} and \cite{sauret2012}.  
It has also been observed in a librating sphere by   \cite{noir2009}. 
Shear instability is present  in planar Stokes layers \cite[e.g.][]{Davis1976} but no clear evidence has been provided so far in cylindrical and spherical geometries. 
In the present study, the forcing amplitude is assumed sufficiently small such that the flow generated at the libration frequency remains stable. 
To fully resolve this leading order flow, we also assume that the libration frequency is larger than twice the mean rotation rate (spin frequency). 
This hypothesis guarantees that no inertial waves are excited and no critical latitude singularity is present.  It will allow us to obtain an expression for the mean zonal 
flow in the whole shell. 
An expression for the mean  zonal flow will also be provided for smaller frequencies (smaller than twice the mean rotation frequency) but it will exhibit a divergence 
associated with the presence of a critical latitude. 
 We shall see  that for any  libration frequency, the presence of an inner core significantly modifies the mean zonal flow and generates  so-called Stewartson layers  
 \cite[][]{stewartson1966} across the cylinder tangent to the inner core.

 The paper is organized as follows.
In section 2, we present  the mathematical framework of the problem. We provide the governing equations and the hypothesis for the asymptotic analysis. 
The leading order solution at the libration amplitude $\eps$ is derived in  section 3. 
 In section 4,  we focus on the mean flow correction at the order $\eps ^2$.  We explain how the nonlinear interactions within the Ekman layer generate a mean zonal flow in the bulk of 
 the shell. We show that a discontinuity appears across the cylinder tangent to the inner core which requires the introduction of viscous layers as for differentially rotating concentric spheres
 \cite[][]{stewartson1966}. The structure of the solution in these layers is provided. We show that both axial vorticity and axial flow are the strongest in these layers. 
 The details of the calculation leading to the expression of the different fields are given in the appendix.  In section 5, the asymptotic solution is compared to numerical results obtained for both a sphere and a shell and a fairly good agreement is obtained.  A brief conclusion with some perspectives is provided in section 6.


\section{Framework of the asymptotic analysis}


We consider a homogeneous and incompressible fluid of kinematic viscosity $\nu$  contained in a spherical shell of external radius $R_{ext}$ and inner radius $R_{int}$. This shell rotates in the laboratory frame at angular velocity which sinusoidally oscillates around a mean angular value $\Omega_0$. We use $R_{ext}$ and ${\Omega_0}^{-1}$ as the length scale and the time scale respectively. 
As both the libration forcing and the geometry are axisymmetric, 
we shall assume that the flow remains axisymmetric, that is its velocity and pressure field are independent of the azimuthal variable $\phi$. It is then convenient to use the spherical coordinates 
$(r,\theta,\phi)$ and to express the velocity field  $(u_r,u_\theta,u_\phi)$ as  functions of $\psi$ and $\chi$ defined by:
\be
u_r  =  \frac{1}{r^2\,\sin \theta}\,\frac{\partial \psi}{\partial \theta}, \qquad u_\theta  =  -\frac{1}{r\,\sin \theta}\,\frac{\partial \psi}{\partial r}, \qquad u_\phi=\frac{\chi}{r\,\sin \theta}.
\label{exp:u}
\ee
The use of a toroidal/poloidal decomposition of the velocity field is also possible \cite[see for instance][]{dormy1998}, but we have preferred  the formulation of 
\cite{stewartson1966} to use the similarity with his analysis. 

In the frame rotating at the mean angular velocity,  the dimensionless Navier-Stokes and continuity equations can be expressed as
\bea
\frac{\partial D^2\psi}{\partial t}   - 2\left(\frac{\sin \theta}{r}\frac{\partial \chi}{\partial \theta}- \cos \theta \,\frac{\partial \chi}{\partial r} \right)  - E\,D^4\psi   
= \frac{2\,\chi}{r^2\,\sin^2\theta}\left(\frac{\sin \theta}{r}\frac{\partial \chi}{\partial \theta}  
  - \cos \theta\,\frac{\partial \chi}{\partial r}\right)  \nonumber \\ 
+\frac{1}{r^2\,\sin\theta}\left(\frac{\partial \psi}{\partial r}\frac{\partial D^2\psi}{\partial \theta}-\frac{\partial \psi}{\partial \theta}\frac{\partial D^2\psi}{\partial r}\right) 
+ \frac{2\,D^2\psi}{r^2\,\sin^2\theta} \left(\frac{\sin \theta}{r}\frac{\partial \psi}{\partial \theta}- \cos \theta\frac{\partial \psi}{\partial r} \right),  \quad \label{motion1}
\eea
\begin{equation}
\frac{\partial \chi}{\partial t} +2 \left( \frac{ \sin \theta}{r}\frac{\partial \psi}{\partial \theta} -\cos\theta \,\frac{\partial \psi}{\partial r} \right) -E\,D^2\chi  =  \frac{1}{r^2\,\sin\theta}\left(\frac{\partial \psi}{\partial r}\frac{\partial \chi}{\partial \theta}-\frac{\partial \psi}{\partial \theta}\frac{\partial \chi}{\partial r}\right),  \label{motion2}
\end{equation}
\noindent where
\begin{equation}
D^2=\frac{\partial^2}{\partial\,r^2}+\frac{\sin \theta}{r^2}\,\frac{\partial}{\partial \theta}\left(\frac{1}{\sin\theta}\,\frac{\partial}{\partial \theta}\right),
\end{equation}
\noindent and $E$ is the Ekman number defined by:
\begin{equation}
E=\frac{\nu}{\Omega\,{R_{ext}}^2} . 
\label{ekman}
\end{equation}

We consider no-slip boundary conditions for a general configuration where the outer sphere and the inner core do not librate at the same amplitude. The boundary conditions in the rotating frame of reference  then read for the functions $\psi$ and $\chi$: 
\bsea 
   \frac{\partial \psi}{\partial r}=\psi=0 \qquad \text{and} \qquad \chi=\eps\,\sin^2\theta\,\cos(\omega t) \qquad \text{at} \qquad r=1 \\
   \frac{\partial \psi}{\partial r}=\psi=0 \qquad \text{and} \qquad \chi=\eps\,\alpha\, a^2\sin^2\theta\,\cos(\omega t) \qquad \text{at} \qquad r=a  \label{inner1}
\esea
where $a=R_{int}/R_{ext}$ is the aspect ratio of the shell, $\eps$ the amplitude of libration of the outer sphere and $\omega$ the frequency of libration. The parameter $\alpha$ 
measures the relative amplitude of libration of the inner core.  For a non-librating inner core, we impose $\alpha=0$, whereas when it is librating at the same amplitude as the outer sphere, $\alpha=1$.  
For the case of a sphere, without inner core, only the condition at $r=1$ has to be applied.

The objective of our work is to characterize the steady flow induced by the librational forcing. 
We assume that the forcing amplitude $\eps$ is small such that the flow remains stable 
with respect to the centrifugal instability. 
Moreover, we consider the limit $\eps \ll 1$ in order to use  asymptotic methods. 
In this limit,  we can expand the functions $\psi$ and $\chi$ in power of $\eps$: 
\bsea
\psi=\eps\,\psi_1+\eps^2\,\psi_2+o(\eps^2) ,\\
\chi= \eps\,\chi_1+\eps^2\,\chi_2+o(\eps^2) .
\label{expansion}
\esea
The first order solution is expected to be the linear response at the librating frequency  forced by the boundary while the second order 
represents the solution generated by the nonlinear interactions of the first order solution with itself. 
As in all weakly nonlinear analysis, both harmonic corrections oscillating at $2 \omega$ and mean flow corrections (the zonal flow) are 
expected. It is this second part owing to its peculiar structure that will be our point of focus.  

\section{Linear solution}

In this section, we consider the solution obtained at the order $\eps$. 
We are interested in the solution for very small values of $E$, as encountered in geophysical applications. For this reason, it is natural to consider $E$ as an
asymptotically small parameter. Nevertheless, this limit is taken after having considered $\eps \rightarrow 0$. 
In the small $E$ limit, we expect viscous effects to be localized in space. In our analysis, we shall require
that, for the first order solution, they are localized near the boundaries. 
This strong constraint can be satisfied if no inertial waves are excited in the bulk by the harmonic forcing. This is verified 
if the frequency of the forcing is outside the range of inertial wave frequencies, that is $\omega >2$, which is the condition that we shall
assume in the following. 

For $\omega >2$, the linear response is expected to be mainly localized close to the boundaries. This solution can be obtained
by performing a classical asymptotic analysis as done for Stokes or  Ekman layers. 
Let us consider first the boundary layer on the outer sphere at $r=1$. 
The boundary layer solution is obtained by introducing the local variable $\tilde{r}=(1-r)/\sqrt{E}$.
Let us write the leading order expression for $\psi$ and $\chi$ as
\bsea  
\psi_1\sim \sqrt{E}\,\psi_1^{(o)} e^{i\om t} + c.c. \\
\chi_1\sim \chi_1^{(o)} e^{i\om t} + c.c. 
\label{exp:psi1}
\esea
\noindent where the superscript $(o)$ indicates that we consider the function in the viscous layer at the boundary of the outer sphere and $c.c.$ denotes complex conjugate. 
Inserting (\ref{expansion}a,b) with (\ref{exp:psi1}a,b) in (\ref{motion1}) and (\ref{motion2}), we obtain, at leading order in $E$ and $\eps$, the equations 
\bsea 
\text{i} \omega \frac{\partial^2 {\psi_1^{(o)}}}{\partial \tilde{r}^2}-2\,\cos \theta\,\frac{\partial {\chi_1^{(o)}}}{\partial \tilde{r}}=\frac{\partial^4 {\psi_1^{(o)}}}{\partial \tilde{r}^4}, 
\label{epsil1}  \\
\text{i} \omega \chi_1^{(o)}+2\,\cos \theta\,\frac{\partial {\psi_1^{(o)}}}{\partial \tilde{r}}=\frac{\partial^2 {\chi_1^{(o)}}}{\partial \tilde{r}^2}, \label{epsil2}
\esea
with the no-slip boundary conditions 
\bsea \psi_1^{(o)} (\tilde{r}=0,\theta) = 0 ~,\\
 \partial_{\tr} \psi_1^{(o)}(\tilde{r}=0,\theta) =0~, \\
 \chi_1^{(o)} (\tilde{r}=0,\theta)= (\sin ^2\theta)/2 ~.
 \esea 
These equations together with the boundary conditions give
 \bsea
\psi_1^{(o)}  =\frac{\text{i}\,\sin^2\,\theta}{4}\left[\frac{1-\text{e}^{-\lambda_+\,\tilde{r}}}{\lambda_+}-\frac{1-\text{e}^{-\lambda_-\,\tilde{r}}}{\lambda_-}\right], \\
\chi_1^{(o)}=\frac{\sin^2\,\theta}{4}\left[\text{e}^{-\lambda_+\,\tilde{r}}+\text{e}^{-\lambda_-\,\tilde{r}}\right],
\esea
where we have defined  
\be
\lambda_{\pm}  =  (1+\text{i})\sqrt{\frac{\omega}{2}\pm \cos\theta} . 
\label{exp:lambda}
\ee
This expression can also be found in \cite{busse1968}. 
Note that the condition $\omega >2$ guarantees that the boundary layer solution does not diverge for any value of $\theta$. 
Such a divergence, called critical latitude singularity, is known to be a source of inner shear layers as documented for the case of precession \cite[see for instance][]{Hollerbach1995}.

A similar solution is obtained in the viscous layer near the inner sphere using the  local variable \mbox{$ \hr = (r-a)/\sqrt{E}$}:
 \bsea
\psi_1^{(i)}  = -\alpha \, \frac{\text{i}\,a^2\, \sin^2\,\theta}{4}\left[\frac{1-\text{e}^{-\lambda_+\,\hr}}{\lambda_+}-\frac{1-\text{e}^{-\lambda_-\,\hr}}{\lambda_-}\right], \\
\chi_1^{(i)}= \alpha\,\frac{a^2\, \sin^2\,\theta}{4}\left[\text{e}^{-\lambda_+\,   \hr}+\text{e}^{-\lambda_-\,   \hr}\right],
\esea

As $  \hr$ (resp. $\tr$) goes to infinity, $\chi_1^{(i)}$ (resp. $\chi^{(o)}$) goes to zero while $\psi_1^{(i)}$ (resp. $\psi^{(o)}$) goes to a non-zero constant.
This part of the solution is responsible for a contribution in the bulk at the order $\eps\,\sqrt{E}$ which corresponds to the Ekman pumping.

\section{Mean zonal flow}

\subsection{Flow asymptotic structure}

We now consider the nonlinear solution at  order $\eps^2$. 
This second order correction is created by the nonlinear interaction of the first order solution with itself. 
Note first that the nonlinear interaction of the solution in the bulk (the Ekman pumping) is expected to generate a correction 
at  order $\eps ^2\,E$. 
As we shall see, this order is actually smaller that the order of the mean flow correction that we will find in the limit $E \to 0$. 
The main contribution in the bulk will be found to come from the nonlinear interactions in the boundary layers
and to be of the form:
\bsea
\psi _2 \sim  \sqrt{E} \,\bpsi_2 (r,\theta) \\
\chi _2 \sim \bchi_2 (r,\theta)
\label{exp:bulk}
\esea
where $\bpsi_2$ and $\bchi_2$ are both independent of time and $E$.

Equations (\ref{motion1}) and (\ref{motion2})  imply that $\bpsi_2$ and $\bchi_2$  would satisfy in the bulk
\be
\left( \cos \theta \,\pa{}{r} - \frac{\sin\theta}{r}\pa{}{\theta}\right) \bpsi_2=0 \quad \text{and} \quad \left( \cos \theta\, \pa{}{r} - \frac{\sin\theta}{r}\pa{}{\theta} \right)\bchi_2=0,
\ee 
\noindent which can be rewritten in cylindrico-polar coordinates $(\rho,\phi,z)$:
\be
\pa{\bpsi_2}{z}=0 \quad \text{and} \quad \pa{\bchi_2}{z}=0.
\ee 
This means that, in the bulk, these two functions would only be dependent of 
the radial coordinate $\rho=r\sin\theta$ and could be written as
\bsea
\bpsi _2 (r,\theta) = \Psi_2 (\rho), \\
\bchi_2(r,\theta) = \Bchi _2(\rho) .
\esea 
This property is nothing but a consequence of the Taylor-Proudman theorem. 

\begin{center}
\begin{figure}
\begin{center}\includegraphics[width=6cm]{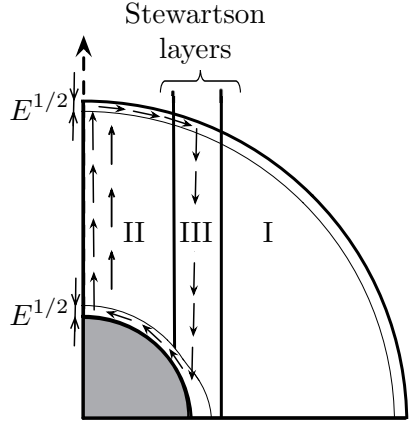}\end{center}
\caption{Structure of the flow in the librating shell showing the presence of three different zones.}
\label{fig:syst_noyau}
\end{figure}
\end{center}

In the next section, we calculate the functions $\Psi_2(\rho)$ and $\Bchi_2(\rho)$ by performing the matched asymptotic analysis of 
the solutions in the boundary layers and in the bulk. 
The asymptotic structure of the problem is provided in figure \ref{fig:syst_noyau}. 
In addition to the boundary layers, the volume is split in two main regions I and II. 
Because one of the boundary layer disappears when we cross the cylinder tangent to the inner core, different approximations are obtained in regions I and II. 
We shall see that the functions are discontinuous across the boundary between regions I and II and that smoothing this singularity requires thin nested layers (region III) in which viscous effects have to be introduced.

In the following, we only consider mean flow quantities: the bars  are then dropped out (e.g. $\bchi _2$ will be noted $\chi_2$).

\subsection{Boundary layer solution near the outer boundary $r=1$}

 Using the boundary layer variable $\tilde{r}=(1-r)/\sqrt{E}$, the equations (\ref{motion1}) and (\ref{motion2}) for the stationary component in the boundary layer at order $\eps^2$ write
\begin{eqnarray}
\frac{\partial^4 {\psi}_2^{(o)}}{\partial \tilde{r}^4}+2\,\cos \theta\,\frac{\partial {\chi}_2^{(o)}}{\partial \tilde{r}} & = & N\!L_1^{(o)}, \label{eq_NL1} \\
\frac{\partial^2 {\chi}_2^{(o)}}{\partial \tilde{r}^2}-2\,\cos \theta\,\frac{\partial {\psi}_2^{(o)}}{\partial \tilde{r}} & = & N\!L_2^{(o)}, \label{eq_NL2}
\end{eqnarray}

\noindent where $N\!L_1^{(o)}$ and $N\!L_2^{(o)}$ are the stationary part of the nonlinear forcing terms given by:
\begin{eqnarray}
N\!L_1^{(o)} & =& -\frac{2\,\cos \theta}{\sin^2\theta}\left[\frac{\partial^2{\psi}_1^{(o)}}{\partial \tilde{r}^2}\,\frac{\partial {\psi}_1^{(o)}}{\partial \tilde{r}}+{\chi}_1^{(o)}\,\frac{\partial {\chi}_1^{(o)}}{\partial \tilde{r}}\right]+\frac{1}{\sin \theta}\left[\frac{\partial {\psi}_1^{(o)}}{\partial \tilde{r}}\frac{\partial^3 {\psi}_1^{(o)}}{\partial \theta \partial \tilde{r}^2}-\frac{\partial {\psi}_1^{(o)}}{\partial \theta}\frac{\partial^3 {\psi}_1^{(o)}}{\partial \tilde{r}^3}\right] \nonumber \\
\label{NL1} \\
N\!L_2^{(o)} & = & \frac{1}{\sin\theta}\left[\frac{\partial {\psi}_1^{(o)}}{\partial \tilde{r}}\frac{\partial {\chi}_1^{(o)}}{\partial \theta}-\frac{\partial {\psi}_1^{(o)}}{\partial \theta}\frac{\partial {\chi}_1^{(o)}}{\partial \tilde{r}}\right] \label{NL2}
\end{eqnarray}

From  (\ref{eq_NL1}) and (\ref{eq_NL2}), we obtain a fifth-order equation for $\psi_2^{(o)}$:
  \begin{eqnarray}
\frac{\partial}{\partial \tilde{r}}\left[\frac{\partial^4}{\partial \tilde{r}^4}+4\,\cos^2\theta\right]\psi_2^{(o)} = \frac{\partial N\!L_1^{(o)}}{\partial \tilde{r}} -2\,\cos\theta\,N\!L_2^{(o)}= N\!L^{(o)} \label{eq2ext}
\end{eqnarray}

This equation has to be solved with the boundary conditions:
\begin{equation}
\psi_2^{(o)}(\tilde{r}=0)=0, \qquad \frac{\partial \psi_2^{(o)}}{\partial \tilde{r}}(\tilde{r}=0)=0 ,  \qquad  \lim\limits_{\tilde{r} \to +\infty} \psi_2^{(o)}(\tilde{r})=\Psi_2(\sin\theta),
\label{CL2ext}
\end{equation}


The expressions of  $N\!L^{(o)}$ and $N\!L_1^{(o)}$ are cumbersome. We can write them in a compact form as 
\bsea
N\!L^{(o)}  = \sum_{l=1}^5\,\left[A_{l}(\theta)+\tr\,B_{l}(\theta)\right]\,e^{-{\mu_l}\,\tr}+c.c. \\
N\!L_1^{(o)}  = \sum_{l=1}^5\,\left[C_{l}(\theta)+\tr\,D_{l}(\theta)\right]\,e^{-{\mu_l}\,\tr}+c.c. 
\label{exp:NL}
 \esea 
 where the exponents $\mu_l$ are related to $\lambda_{\pm}$ defined in (\ref{exp:lambda}) by
 \be
  \mu_1  = \lambda_+  +\lambda_-^*, \quad \mu_2 =  \lambda_++\lambda_+^*, \quad \mu_3= \lambda_-+\lambda_-^*,\quad \mu_4  =  \lambda_+ , \quad \mu_5= \lambda_-.
 \ee
Expressions for  the coefficients $A_l(\theta)$, $B_l(\theta)$ , $C_l(\theta)$  and $D_l(\theta)$ are provided in appendix~A. 
 
The solution to  (\ref{eq2ext}) which satisfies the boundary conditions (\ref{CL2ext}) reads
\be
\psi_2^{(o)}(\tr,\theta)  =  \phi_2^{(o)} (\tr,\theta) +\Psi_2(\sin\theta)\left[1-\frac{\kappa^*\,\text{e}^{-\kappa\tr}-\kappa\,\text{e}^{-\kappa^*\tr}}{\kappa^*-\kappa}\right] \label{solutionBL11}
\ee
where 
\be
\ba{l}
\phi_2^{(o)}(\tr,\theta) =  {\displaystyle \sum_{l=1}^5 \left\{ -\fracd{1}{4\cos^2\theta+{\mu_l}^4}\Biggl[\frac{B_{l}+{\mu_l}A_{l}}{{\mu_l}^2}+\frac{4\,{\mu_l}^2\,B_{l}}{4\cos^2\theta+{\mu_l}^4}\Biggr]\Biggl[\text{e}^{-{\mu_l}\tr}+\frac{{\mu_l}-\kappa^*}{\kappa^*-\kappa}\,\text{e}^{-\kappa\tr}  \right.  }\\  
 \left. \qquad +\fracd{{\mu_l}-\kappa}{\kappa-\kappa^*}\,\text{e}^{-\kappa^*\tr}\Biggr]-\frac{B_{l}}{{\mu_l}(4\cos^2\theta+{\mu_l}^4)}\Biggl[\tr\,\text{e}^{-{\mu_l}\tr}-\frac{\text{e}^{-\kappa\tr}-\text{e}^{-\kappa^*\tr}}{\kappa^*-\kappa}\Biggr]   \right\}  + c.c. 
\ea
\label{exp:bphi(o)}
\ee
and where we have defined
\begin{equation}
\kappa=(1+\text{i})\sqrt{\cos\theta}. 
\end{equation}

From  (\ref{eq_NL1}),  we  obtain $\chi_2^{(o)}$, which must satisfy the matching condition:
\be
\lim\limits_{\tilde{r} \to +\infty} \chi_2^{(o)}(\tilde{r})=\Bchi_2(\sin\theta),
\ee
as
 \be
 \chi_2^{(o)}(\tr,\theta)=-\frac{1}{2\,\cos\theta}\Biggl[\frac{\partial^3\psi_2^{(o)}}{\partial\tr^3}+\left\{\sum_{l=1}^5 \left(\frac{D_{l}+{\mu_l}\,C_{l}}{{\mu_l}^2}+\frac{\tr\,D_{l}}{{\mu_l}}\right)\text{e}^{-{\mu_l}\tr}  +c.c. \right\} \Biggr]+\Bchi_2(\sin\theta) .~~~~~~~ \label{solutionBL}
 \ee 
  The above expression satisfies the no-slip boundary condition  $\chi_2^{(o)}(0,\theta)=0$ only if  $\Psi_2$ and $\Bchi_2$ satisfies the 
 relation:
 \be
 \Bchi_2(\sin\theta ) =  \BF(\sin\theta;\om) - 2 \sqrt{\cos\theta} \, \Psi_2(\sin\theta) 
 \label{relation(o)}
 \ee
 where 
 \be
 \BF(\sin\theta;\om) =  \frac{1}{2\,\cos\theta}\Biggl[\left.\frac{\partial^3\phi_2^{(o)}}{\partial\tr^3}\right|_{\tr=0} + \left\{ \sum_{l=1}^5 \frac{D_{l}+{\mu_l}\,C_{l}}{{\mu_l}^2}  +c.c. \right\}\Biggr]
 \ee
 
 Equation (\ref{relation(o)}) is a constraint obtained from the matching between the solution in the bulk
 and the solution in the boundary layer on the outer sphere. 
  
 A similar constraint is expected from the boundary layer on the inner sphere.  
 The same analysis in this boundary layer leads to
 \be
 \Bchi_2(a\sin\theta ) =  \alpha^2 a^2 \BF(\sin\theta;\om) + 2 \sqrt{\cos\theta} \, \Psi_2(a\sin\theta) .
 \label{relation(i)}
 \ee


 \subsection{Solution in the bulk}
 
 Relations (\ref{relation(o)}) and (\ref{relation(i)})  permit us to obtain the functions
 $\Psi_2$ and $\Bchi_2$ which characterize the mean flow correction at leading order in the bulk.
 
 We first consider region I, defined by $\rho> a$ (see figure \ref{fig:syst_noyau}). In this region, the 
 only contribution comes from the viscous layer close to the outer sphere.   
 However the flow in this region must satisfy a condition of symmetry on the equator: the axial flow should be antisymmetric with respect to the equator 
 and therefore should vanish
 on the equator. Since it does not depend on the axial coordinate, this condition implies  that it should vanish everywhere in region I. 
 In other words
 \be 
 \Psi_2(\rho) =0,
 \label{exp:Psi2I}
\ee
from which we deduce from (\ref{relation(o)}), that for $\rho>a$ (i.e. in region I):
\be
\Bchi_2(\rho)= \BF(\rho;\om).
\label{exp:Bchi2I}
\ee

In region II, both the conditions of matching with the inner and outer boundary layers have to be considered.
Replacing $\sin\theta$ by $\rho$ in (\ref{relation(o)}) and $a\sin\theta$ by $\rho$ in (\ref{relation(i)}), lead to
\bsea 
\Bchi_2(\rho) & = &  - 2 \,\left[1 - \rho^2\right]^{1/4}\, \Psi _2(\rho)  + \BF(\rho;\om),\\
\Bchi_2(\rho)&  = & 2 \,\left[1 - (\rho/a)^2\right]^{1/4}\, \Psi _2(\rho) +\alpha^2 \,a^2 \, \BF(\rho/a;\om) ,
\esea
from which we deduce, for $\rho<a$ (in region II)
\bsea 
\Psi_2(\rho)& =&   \frac{\BF(\rho;\om) - \alpha^2  a^2 \BF(\rho/a;\om)}{2(1-\rho^2)^{1/4}+2 \left[1-(\rho/a)^2\right]^{1/4}},\\
\Bchi_2(\rho) &=&  \frac{\alpha^2  a^2 (1-\rho^2)^{1/4}\BF(\rho/a;\om) +\left[1-(\rho/a)^2\right]^{1/4}\BF(\rho;\om)}{(1-\rho^2)^{1/4}+ \left[1-(\rho/a)^2\right]^{1/4}} .
\label{exp:Psi2II}
\esea
 Using (\ref{exp:u}), we can also calculate from $\Psi_2$ and $\Bchi_2$,    the mean azimuthal velocity $u_{\phi _2}$, the mean axial velocity 
 $u_{z_2}$  and the mean axial vorticity $\om_{z_2}$   which satisfy   in regions I and II
 \be
 u_{\phi_2} =  \frac{\Bchi_2}{\rho} ,~~
 u_{z_2} =  \frac{\sqrt{E}}{\rho}\pa{\Psi_2}{\rho}, ~~
\om_{z_2} = \frac{1}{\rho}\pa{\Bchi_2}{\rho}.
 \label{exp:uphi2Bulk}
 \ee

 Note that expressions (\ref{exp:Psi2II}a,b) resemble the expressions obtained by \cite{proudman1956} for the flow between two spheres rotating at different speeds. The difference is in the numerator which is now a more complicated function associated with the 
 nonlinear interaction  in the boundary layer of the flow generated by libration. 
 However, as in \cite{proudman1956}, $\Psi_2$  and $\Bchi_2$ exhibit a discontinuity across $\rho=a$. In particular, note that 
 \bsea
 \Bchi_2(\rho) \simd{\rho}{a^+} \Bchi^+= \BF(a;\om) , \\
  \Bchi_2(\rho) \simd{\rho}{a^-} \Bchi^-=\alpha^2 a^2  \BF(1;\om) = \frac{\alpha^2 a^2}{2 \,\om^2} .
  \esea 
 The singularities of $\Psi_2$ and $\Bchi_2$  can be smoothed
 across a series of viscous layers  as first shown by \cite{stewartson1966}. 
 In the next section, we provide approximations of the solution in these layers. 
 
   \begin{figure}
\begin{center}
\includegraphics[width=8cm]{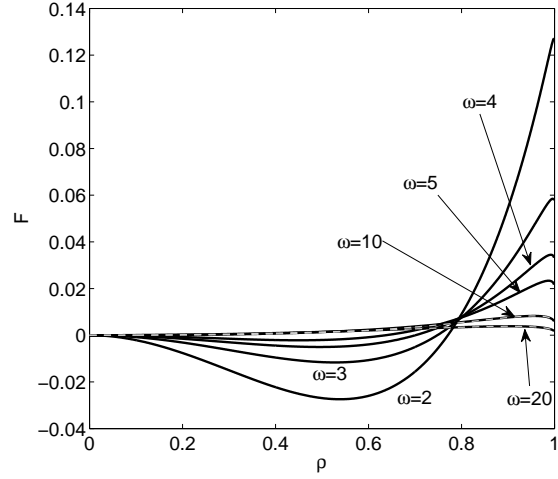}
\end{center}
\caption{The function $\BF(\rho;\om)$ versus $\rho$ for different values of 
the libration frequency $\om$. This function characterizes $\Psi_2$ and $\Bchi_2$ in the bulk. 
Expression (\ref{exp:Finf}) is plotted in dash gray for $\om=10$ and
$\om=20$. }
\label{fig:BF}
\end{figure}
  \begin{figure}
\begin{center}
\includegraphics[width=8cm]{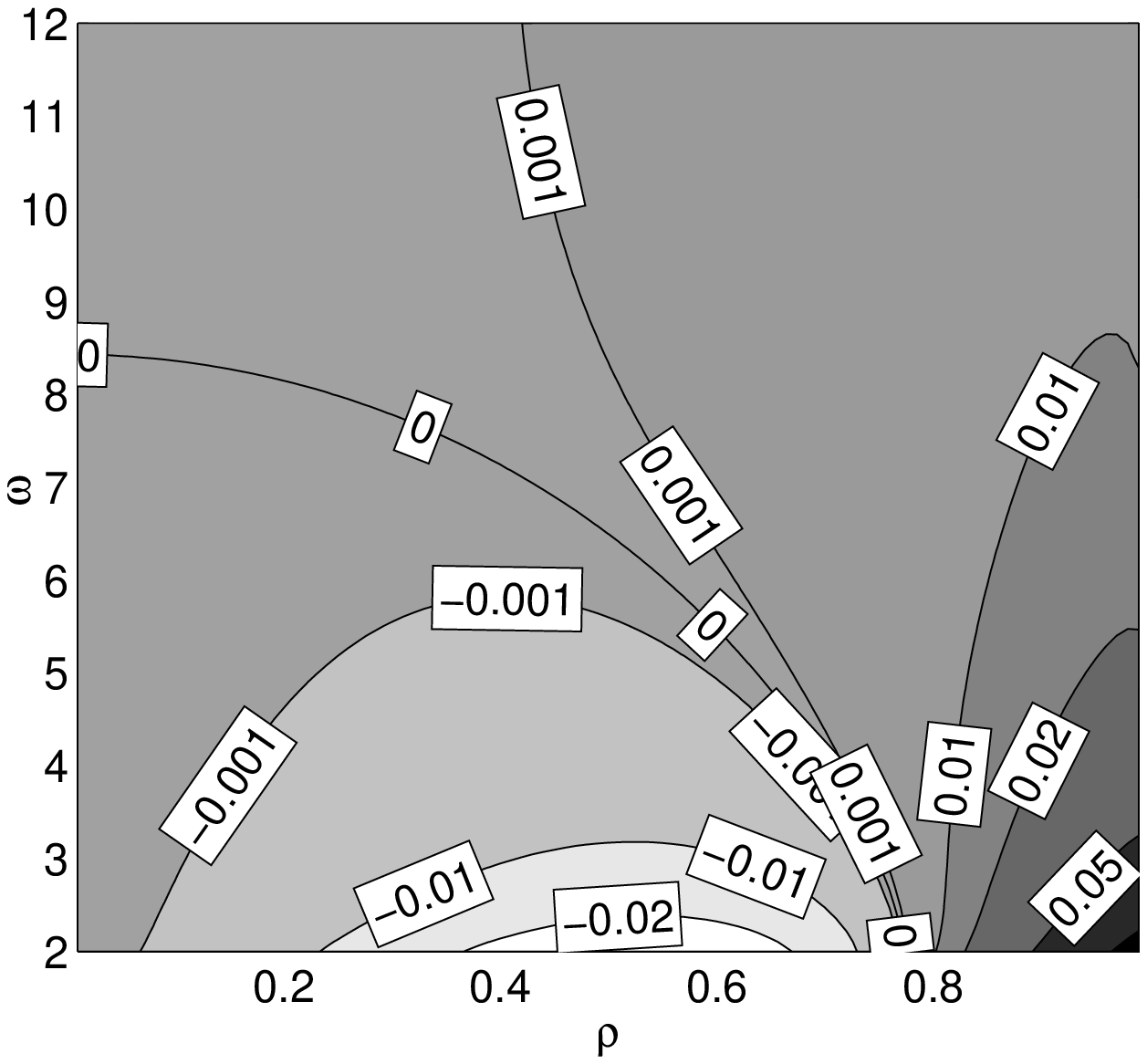}
\end{center}
\caption{Contour levels of the function $\BF(\rho;\om)$ in the ($\rho$, $\om$) plane.}
\label{fig:contourF}
\end{figure}
  It is worth noting that the solution in region II can be deduced for the solution in region I. In other words, if we exclude the Stewartson layers, the solution in a spherical shell
 can be deduced from the solution in a sphere. 
 Indeed, all the information is contained in the function $\BF(\rho;\om)$ which defines $\Bchi_2$ in region I (or in a sphere). Unfortunately, we have not been
 able to obtain a simple expression for this function [see formula (\ref{exp:BF+}) given in appendix B]. In figure \ref{fig:BF}, we have plotted $\BF(\rho;\om)$ versus 
 $\rho$ for various values of the libration frequency $\om$. A contour plot of $\BF(\rho;\om)$ in the $(\rho ,\om )$ plane is also  shown in figure \ref{fig:contourF}.
 Interestingly, we observe that $\BF$ changes sign as $\rho$ varies when $\om < 8.5$. This means that the mean azimuthal flow generated by libration
 could be anticyclonic close to the axis and cyclonic further away.

 For large libration frequency $\om$, an asymptotic expression for $\BF(\rho;\om)$ is obtained as
 \be 
 \BF(\rho;\om) \simd{\om}{\infty} \frac{\rho^2\sqrt{1-\rho^2}}{4\,\om}- \frac{\sqrt{2}\,\rho^2\,(1-\rho^2)^{3/4}}{4\, \om ^{3/2}} + \frac{\rho^2\,(7\rho^2 -5)}{4\, \om^2} + o\left(\frac{1}{\om^2}\right) 
 \label{exp:Finf}
 \ee
 from which we can easily deduce the behavior of $\Bchi_2$, $u_{\phi_2}$, $\Psi_2$ and $u_{z_2}$.
In figure \ref{fig:BF}, we can observe that (\ref{exp:Finf}) provides a very good approximation for $\BF$ for $\om=10$ and larger.

 
Close to $\rho=0$, the function $\BF (\rho;\om)$  can be related to the expression obtained for a cylinder \cite[][]{wang1970}.
We have $\BF (\rho;\om) \sim  \Om_{Cyl}(\om) \rho^2$  as $\rho\rightarrow 0$ which leads to
\be
\Bchi_2 \simd{\rho}{0} \frac{(\alpha^2 +1)}{2}\, \Om_{Cyl}(\om) \,\rho^2 .
\ee
  The function $\Om_{Cyl}(\om)$ corresponds to the mean angular velocity generated by the libration of a cylinder. 
  Its  expression was first obtained by \cite{wang1970} \cite[see also][]{sauret2012}.
 Note  the simple expression for the function  $\BF$ at $\rho=1$:
 $\BF(1;\om) = 1/(4\,\om^2)$.

 It is also interesting to provide expressions for $\Psi_2$ and $\Bchi _2$  in the limit of thin shell ($a\rightarrow 1$):
 \bsea
 \Psi _2 \sim (1-\alpha^2) \frac{\BF(\rho;\om)}{4(1-\rho^2)^{1/4}} + O(1-a) , \\
 \Bchi_2 \sim \frac{1+\alpha^2}{2} \BF(\rho;\om) .
\label{exp:thin}
\esea 
Expression (\ref{exp:thin}a) implies that $\Psi_2$ becomes small of order (1-a) when inner and outer sphere are librating 
at the same amplitude ($\alpha ^2 =1$). Expression (\ref{exp:thin}b) means that the $\rho$ dependence of the function $\Bchi _2$ in the thin shell limit
is the same as for a full sphere.

 \subsection{Solution in the Stewartson layers}
 
\begin{figure}
\begin{center}
\includegraphics[width=6cm]{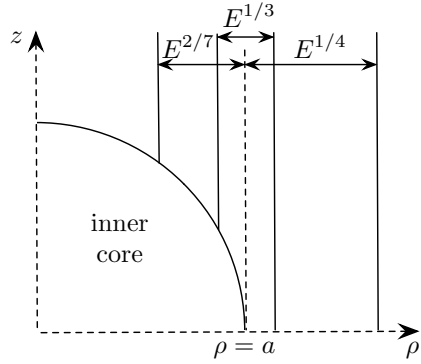}
\end{center}
\caption{Structure of the Stewartson layers. The inner layer has a thickness of $E^{1/3}$ and the external and internal outer layers have a thickness of $E^{1/4}$ and $E^{2/7}$ respectively.}
\label{fig:Stewartson}
\end{figure}

 The structure of the viscous layers is the same as in \cite{stewartson1966} and is sketched in figure \ref{fig:Stewartson}. 
 There are three different layers around $\rho=a$, two outer layers scaling as $|\rho-a|= O(E^{1/4})$ for $\rho>a$ and $|\rho -a |= O(E^{2/7})$ for $\rho<a$, and
 an inner layer where $|\rho-a|=O(E^{1/3})$. 
 As shown by \cite{stewartson1966}, the outer layers guarantee the continuity of $\chi_2$  (and $u _{\phi_2}$) and of its first derivative, while the inner layer is
 for the smoothing of $\psi_2$  (and $u_{z_2}$) and higher derivatives of $\chi_2$.  
 More details on the calculation leading to  approximations of the solution in each layer  are given in  appendix C.  In this section, we only
 provide the leading approximation in each layer. In the appendix, we explain how the next order which is $O(E^{1/28})$ can be obtained.  
 
\begin{figure}
\begin{center}
\includegraphics[width=8cm]{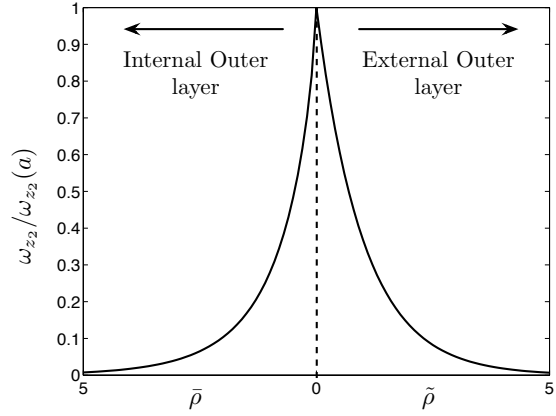}
\end{center}
\caption{Axial vorticity $\om_{z_2}/\om_{z_2}(a)$  as a function of $\brho =(a-\rho)\left[a/(2\,(1-a^2)^2\,E^2)\right]^{1/7}$ for $\rho<a$ (internal outer layer), and
$\trho = (\rho -a)/\left[(1-a^2)^{3/8}\,E^{1/4}\right]$ for $\rho>a$ (external outer layer). }
\label{fig:omegaz}
\end{figure}

 In the outer layers, simple expressions can be obtained. 
 In the external outer layer ($\rho > a$), $\chi_2$ varies exponentially between the two values 
$\Bchi^{\pm}=\lim_{\rho\rightarrow a^{\pm}}\Bchi_2(\rho)$ of  $\Bchi_2$ on either side of $a$ in regions I and II: 
\be
\chi_2 \sim \Bchi^+\left(1- e^{-\trho} \right) +  \Bchi^- e^{-\trho},
\ee
with 
\be
\trho = (\rho -a)/\left[(1-a^2)^{3/8}\,E^{1/4}\right] 
\ee
In the internal outer layer ($\rho<a$), it is a constant
 \be
\chi_2 \sim \Bchi^-.
\ee
The function $\chi_2$ (and therefore $u_{\phi 2}$) is thus of same order in the Stewartson layers than in the bulk.
By contrast, the vorticity associated with the mean flow correction is larger in the outer layers.
Its main component is the axial component $\om_{z_2}$ which  reads
\bsea 
\om _{z_2} \sim E^{-1/4} \frac{\delta\Bchi}{a\,(1-a^2)^{3/8}} \,e^{-\trho} ~~{\rm in ~the ~external~ outer ~layer}, \\
\om _{z_2} \sim E^{-1/4} \frac{\delta\Bchi}{a\,(1-a^2)^{3/8}} \,\frac{\bbf '(\brho)}{\bbf '(0)} ~~{\rm in ~the ~internal~ outer ~layer}, 
\esea
where the internal outer layer variable $\brho$ is defined by
\be
\brho=(a-\rho)\Biggl[\frac{a}{2\,E^2(1-a^2)^2}\Biggr]^{1/7}.
\ee
The function $\bbf$ is obtained in the appendix. It can be expressed in terms of Bessel functions [see expression (\ref{exp:f})]. 
In figure \ref{fig:omegaz}, we have plotted $\om_{z_2}/\om_{z_2}(a)$ in each outer layer.
Note that the vorticity is strongly peaked at  $\rho=a$ where it reaches its maximum value (in amplitude)
\be
\om_{z_2}(a)= E^{-1/4}\frac{\delta\Bchi }{a\,(1-a^2)^{3/8}}.
\ee
The discontinuity of the derivative at this point is smoothed in the inner layer.

 \begin{figure}
\begin{center}
\includegraphics[width=6.7cm]{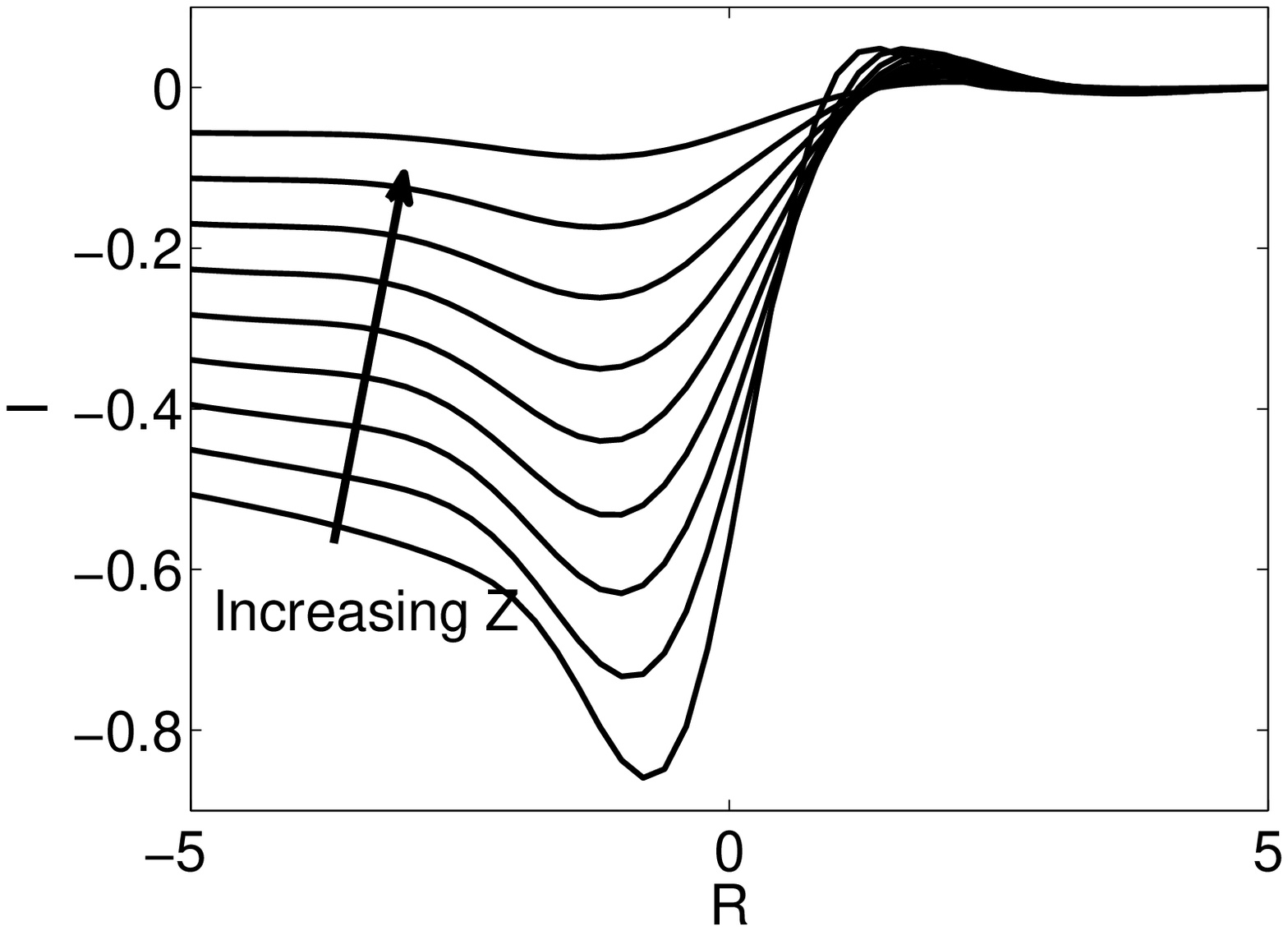}
\includegraphics[width=6.7cm]{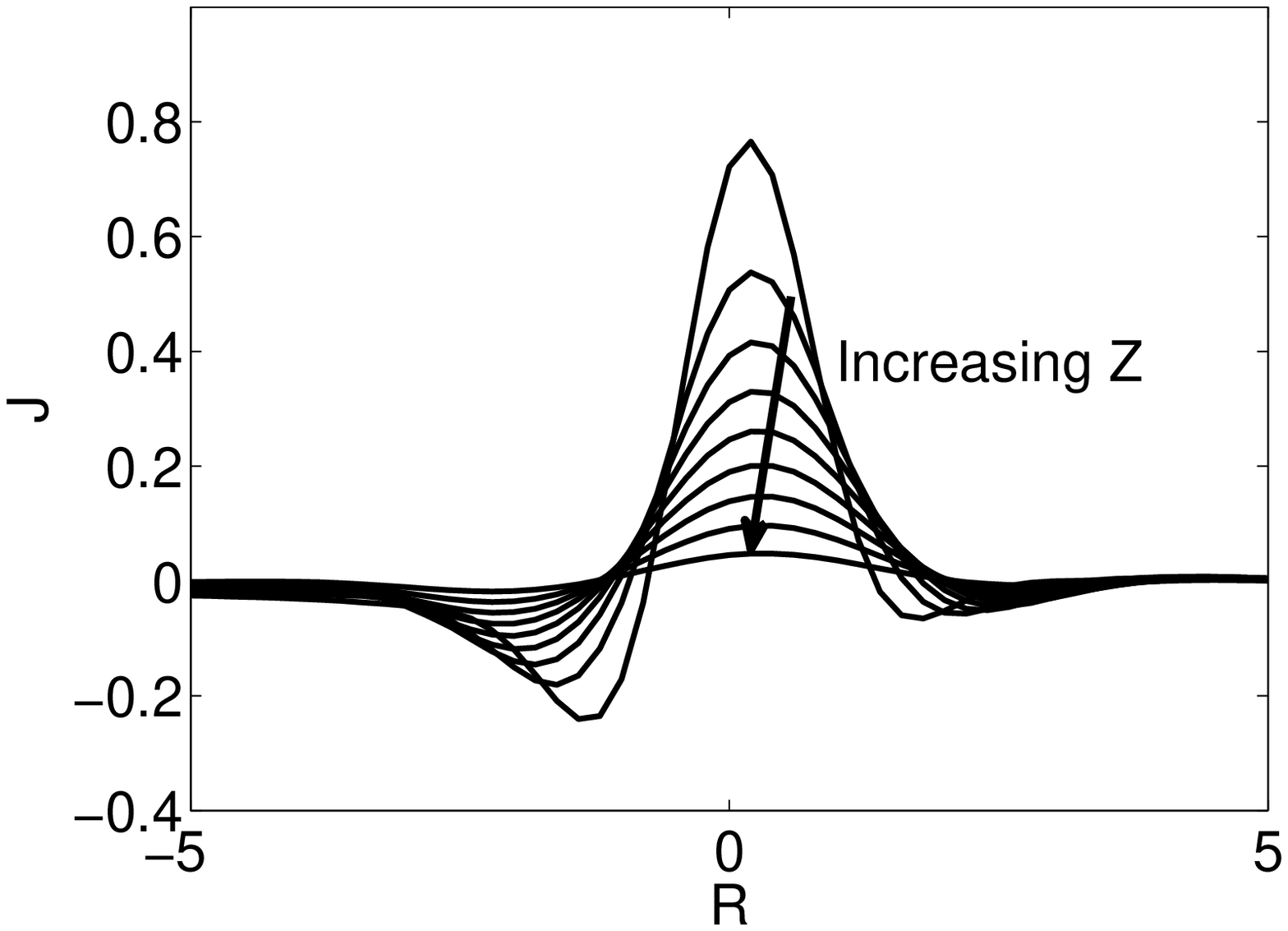}
\\
(a) \hspace{6cm} (b)
\end{center}
\caption{The functions (a) $\bI(Z,R)$ and (b) $\bJ(Z,R)$ versus R for different values of $Z$ ($Z=0.1:0.1:0.9$). 
These two functions provide the spatial variation  in the inner layer, of $\psi_2$ and $u_{z_2}$ respectively.  }
\label{fig:uz}
\end{figure} 
Contrarily to the azimuthal velocity and the axial vorticity, 
the function $\psi_2$ and the axial velocity depend on the axial variable $z$ at leading order. 
This dependence is linear in the outer layers [see expressions (\ref{exp:psi2ExOut}) and (\ref{exp:psi2InOut}) in the appendix C] 
but it is more complex in the inner layer.
It is in this inner layer where $|\rho -a|=O(E^{1/3})$ that $\psi_2$ and $u_{z_2}$ are the largest. 
They scale respectively as $\psi_2=O(E^{19/42})$ and $u_{z_2}=O(E^{5/42})$. Note in particular that 
the axial flow is much larger than in region II  where it is $O(\sqrt{E})$.
Unfortunately, the expressions for $\psi_2$ and  $u_{z_2}$ are not as simple as for $\chi_2$. 
In the inner layer ($|\rho-a|=O(E^{1/3})$), $\psi_2$ and $u_{z_2}$ can
be written as
\bsea
\psi_2 \sim E^{19/42}\Lambda_3(a,\om)\, {\bf I}\left(\frac{z}{\sqrt{1-a^2}},\frac{\rho -a}{E^{1/3}(1-a^2)^{1/6}}\right) , \\
u_{z_2} \sim E^{5/42} \Lambda_4(a,\om)\, {\bf J}\left(\frac{z}{\sqrt{1-a^2}},\frac{\rho -a}{E^{1/3}(1-a^2)^{1/6}}\right) ,
\esea
where  ${\bf I}$ and ${\bf J}$ are the functions given in the appendix by (\ref{exp:I}) and  (\ref{exp:J}) and 
$\Lambda_3$ and $\Lambda_4$ are related to $\delta \Bchi= \Bchi ^+ - \Bchi^-= \BF(a;\om) - \alpha ^2 a^2/(4\,\om^2) $ by
\bsea 
\Lambda_3(a,\om) = \frac{\alpha_o \, a^{3/28} }{(1-a^2)^{11/84}}\,\delta \Bchi (a,\om) , \\
\Lambda_4(a,\om) = \frac{\alpha_o}{a^{25/28}\,(1-a^2)^{25/84}} \,\delta \Bchi (a,\om),
\esea
with $\alpha_o = -1/[2^{6/7}\,\bbf'(0)] \approx 0.45$.

The variations of the functions $\bI(Z,R)$ and $\bJ(Z,R)$ with respect to $R$ at various height $Z$
are shown in figure \ref{fig:uz}. Note that both functions are the largest for small values of $Z$, that is close to the inner sphere.
It is also interesting to note the peculiar form of the axial flow profile, which is composed of a main jet surrounded by two weaker counter-propagating jets.

 \begin{figure}
\begin{center}
\includegraphics[width=6.7cm]{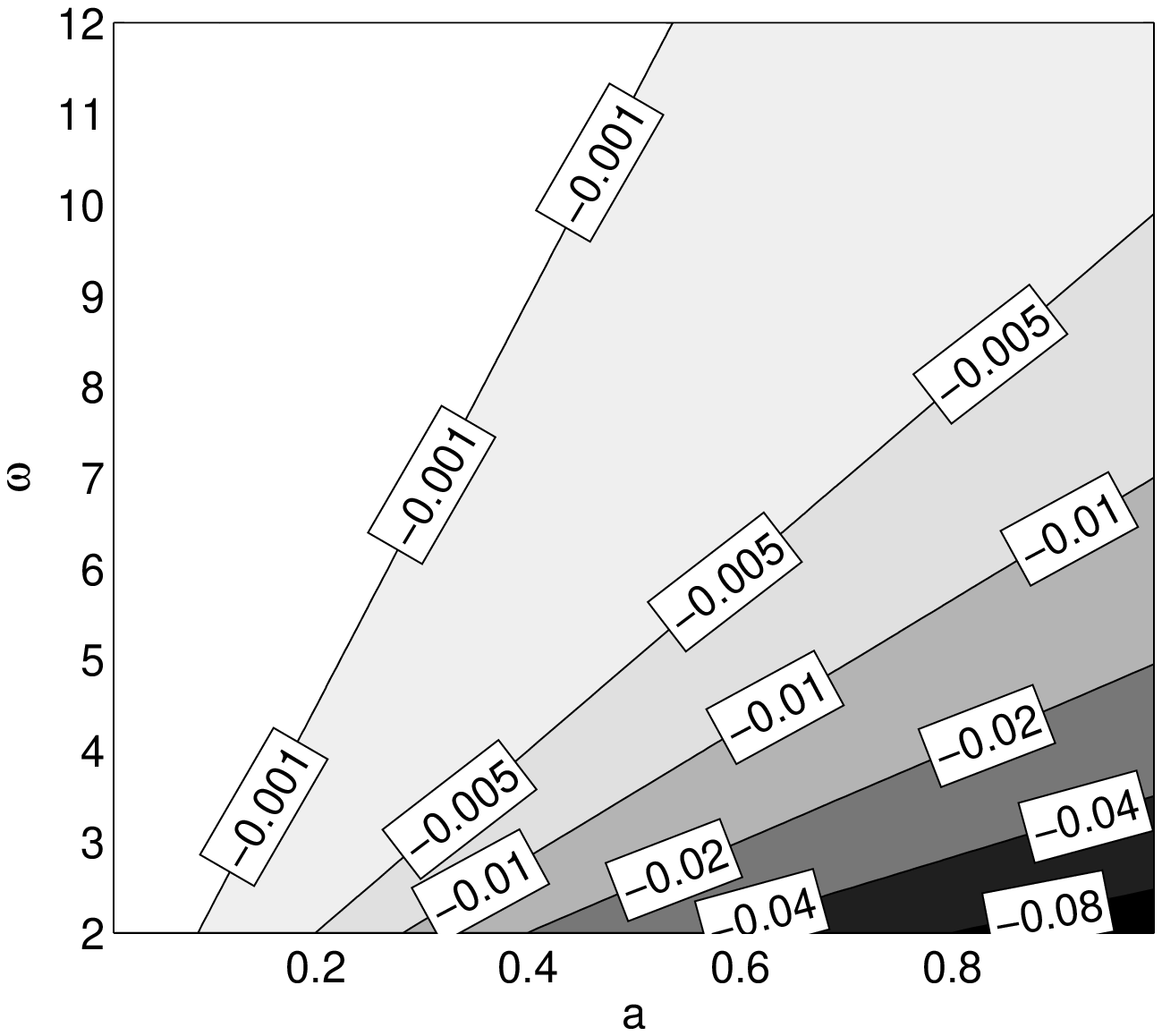}
\includegraphics[width=6.7cm]{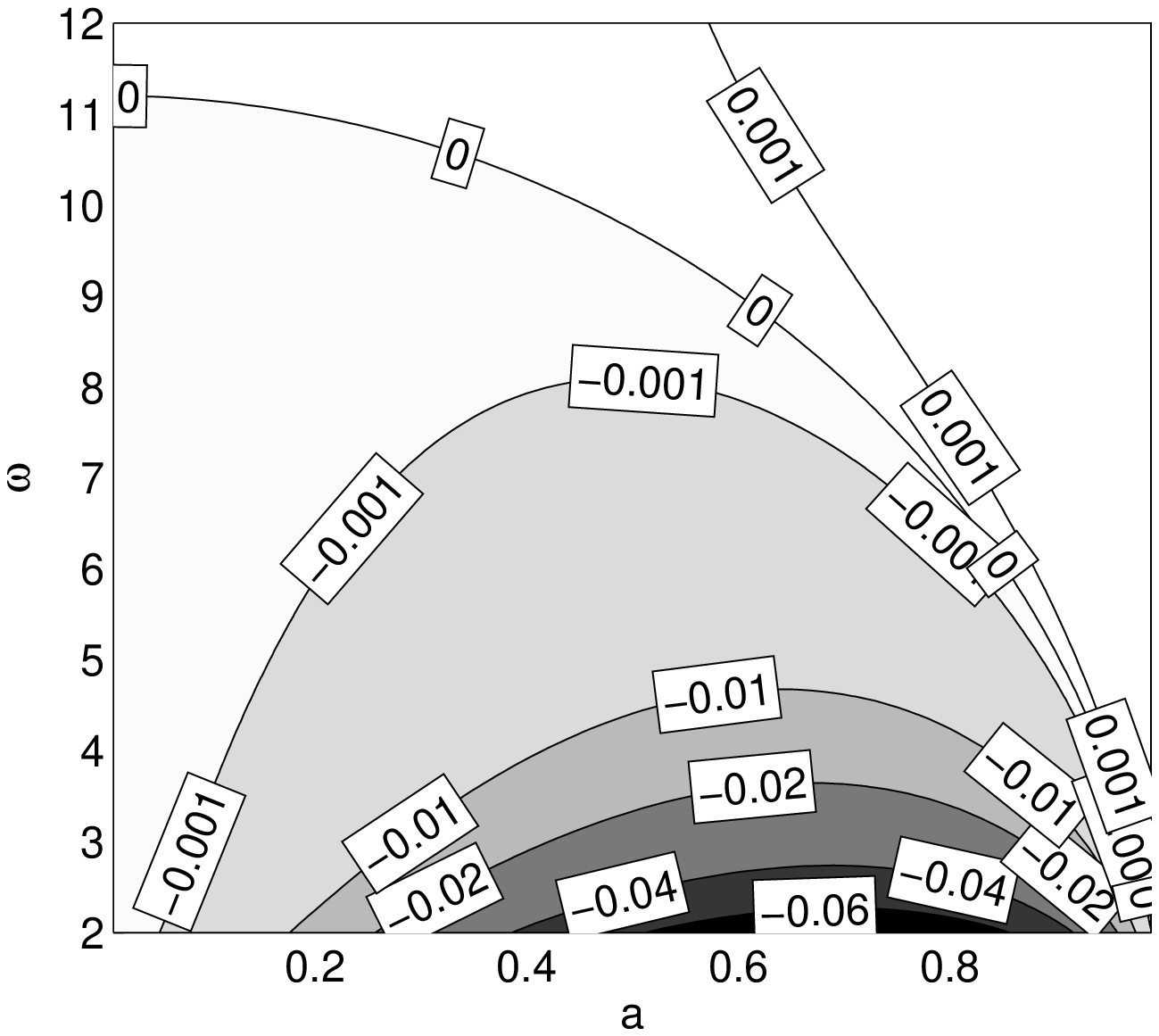} \\
(a) \hspace{6cm} (b)
\end{center}
\caption{Contour of the function $\delta\Bchi(a,\om)$ in the ($a$, $\om$) plane.  (a) when only the inner core is librating: $\delta\Bchi(a,\om)=-a^2/(4\,\om^2)$ . (b)
 when both spheres librate at the same amplitude ($\alpha=1$).  }
\label{fig:DeltaChi}
\end{figure}
It is important to point out that $\psi_2$, $u_{z_2}$ and $\omega_{z_2}$ depend on $\om$ and $\alpha$ only via the function 
$\delta\Bchi$ and that they are all proportional to that function. When the inner core is not librating, $\alpha =0$, and  $\delta\Bchi$ reduces to
the function $\BF(a;\om)$ which has been plotted in figure \ref{fig:contourF}.
The libration of the inner sphere reduces the contribution from the outer sphere  by a simple quantity $\alpha^2 a^2/(4\,\om^2)$. 
If only the inner core was librating (the outer sphere not librating) and if we had used for $\eps$ the amplitude of libration of the inner sphere,
we would have only obtained the inner sphere contribution $\delta\Bchi  = -a^2/(4\,\om^2)$.  
 The contours of this function in the $(a,\om)$ plane  are plotted in figure \ref{fig:DeltaChi}(a).  This plot which describes the contribution
from the inner sphere has to be compared with the one obtained for the outer sphere contribution shown in figure \ref{fig:contourF}. 
We clearly see that the contribution from the inner sphere is always of the same sign whatever $a$ and $\om$: therefore, the libration of
the inner core always generates a negative (anticyclonic) mean axial vorticity  and a main axial flow which goes toward the inner sphere. 
In figure \ref{fig:DeltaChi}(b), we have plotted the variations of $\delta\Bchi$ with respect to both variables
$a$ and $\om$ when inner and outer spheres librate at the same amplitude ($\alpha=1$).
Note that if we compare this plot with figure \ref{fig:contourF}, we observe that the inner sphere provides a non-negligible contribution as soon as the radius  $a$ of the inner core is not small. 
In particular, it strongly increases the domain of parameters where the mean axial vorticity is anticyclonic.


\section{Numerical comparison}

In this section, the results of the asymptotic theory are compared to numerical results obtained for finite values of $\eps$ and $E$.
Both published results obtained by \cite{calkins2010} and  results  obtained 
using the finite elements commercial software Comsol Multiphysics$^{\copyright}$ are considered. 
Both codes rely on a 2D (axisymmetric) model to be able to reach $E$ as small as $E=10^{-6}$.
For more details  about our own numerical procedure, the reader is referred to \cite{sauret2010,sauret2012} where the numerical procedure has already been used 
for similar problems. 

The first and easiest comparison that can be made is for the libration in a sphere. In that case,
there are no Stewartson layers and no axial flow in the bulk. The main contribution to the mean flow correction is an azimuthal flow which is 
theoretically expected to be 
\be
u_{\phi} \sim \eps ^2 \frac{ \BF(\rho;\om)}{\rho}, 
\label{exp:uphiT}
\ee
for small Ekman number E and small libration amplitude $\eps$. 
\begin{center}
\begin{figure}
\begin{center}
\includegraphics[width=9cm]{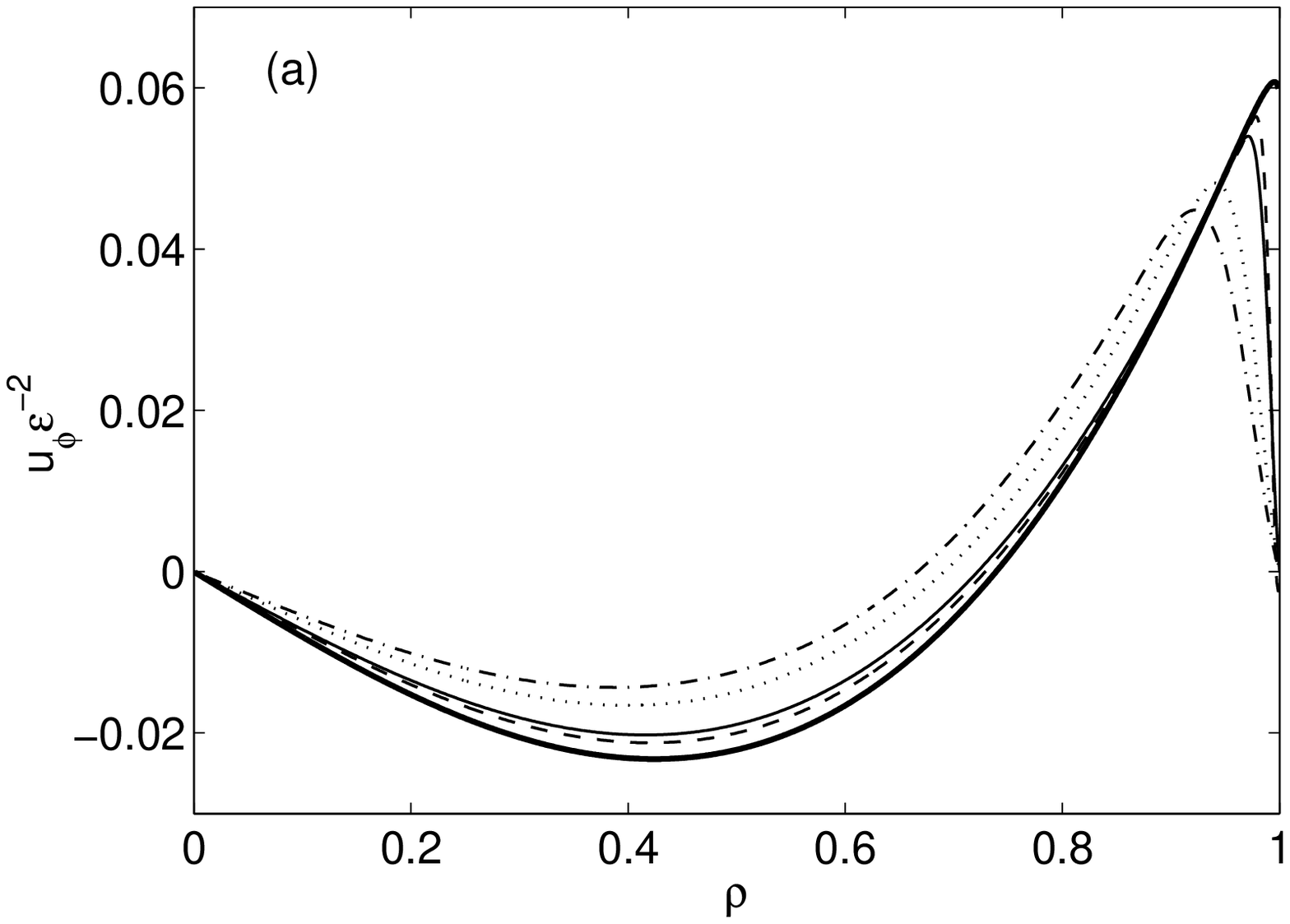}
\includegraphics[width=9cm]{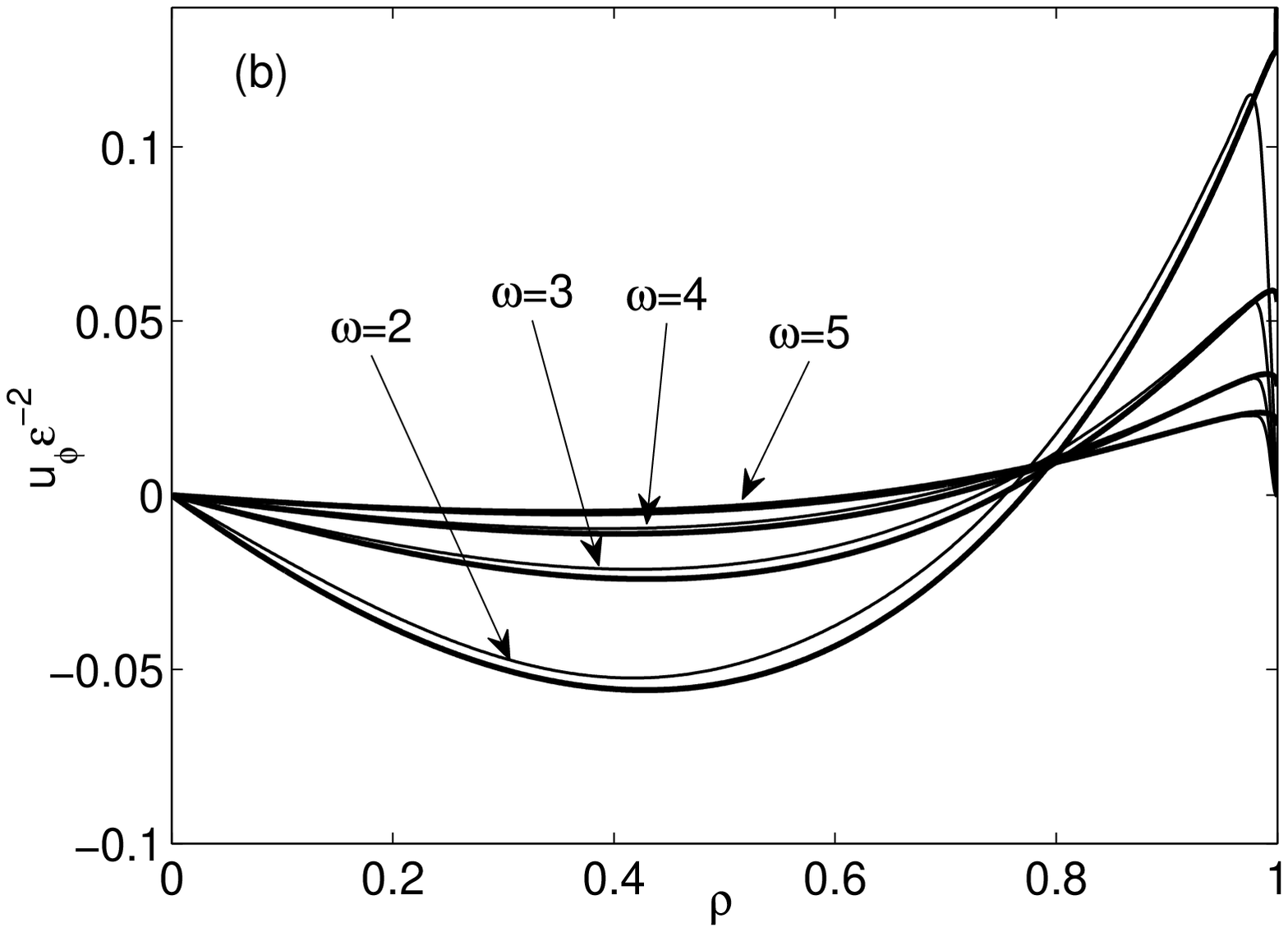}
\end{center}
\caption{
Mean azimuthal flow $u_{\phi_2}=u_{\phi}\eps ^{-2}$ in a sphere. Comparison of numerical results (thin lines) obtained in the equatorial plane ($z=0$) with 
theoretical results (thick line). 
(a)  For fixed frequency and amplitude (here $\omega=3$; $\epsilon=0.1$) and different  Ekman numbers (Dot-dashed line: $E=10^{-3}$; Dotted line: $E=5.10^{-4}$; 
(Thin) solid line: $E=10^{-4}$;  Dashed line: $E=4.10^{-5}$); (b) 
for fixed Ekman number and amplitude ($E=5. 10^{-5}$, $\eps=0.01$) and different frequencies. }
\label{fig:syst_sphere}
\end{figure}
\end{center}
Thus, the theory predicts that the zonal flow should scale as the square of the libration amplitude and be independent of the Ekman number.
These properties 
have been tested by comparing numerical results obtained with Comsol Multiphysics to the theory for  $\eps$ ranging from 0.01 to 0.3 and $E$ ranging from $10^{-3}$ to $4. 10^{-5}$.

In figure  \ref{fig:syst_sphere}(a), we display the results obtained for fixed frequency and amplitude and different Ekman numbers. 
We can observe that the numerical curve which is already close to the theoretical curve for $E=10^{-3}$ gets even closer as the Ekman number decreases. 
The departure obtained close to $\rho=1$ is associated with the boundary layer on the outer sphere which is not included in the theoretical expression (\ref{exp:uphiT}) for 
the solution in the bulk.  When $E$ decreases, the boundary layer which scales as $E^{1/2}$ becomes thiner, as observed. 
The amplitude $\eps$ has also been varied but its effect has been found very weak. For instance,  the curves obtained for $\eps=0.01$ and $\eps=0.3$ cannot be 
distinguished  from those plotted for $\eps=0.1$ in figure  \ref{fig:syst_sphere}(a).

In figure \ref{fig:syst_sphere}(b), we have compared the theory to numerical results for different libration frequencies. Again, we observe a very good agreement for 
all frequencies. 
These results provide a validation of the asymptotic analysis that has led to expression (\ref{exp:uphiT}). 

\begin{center}
\begin{figure}
\begin{center} (a) \hspace{11cm} ~\\[-1cm]
\includegraphics[width=10cm]{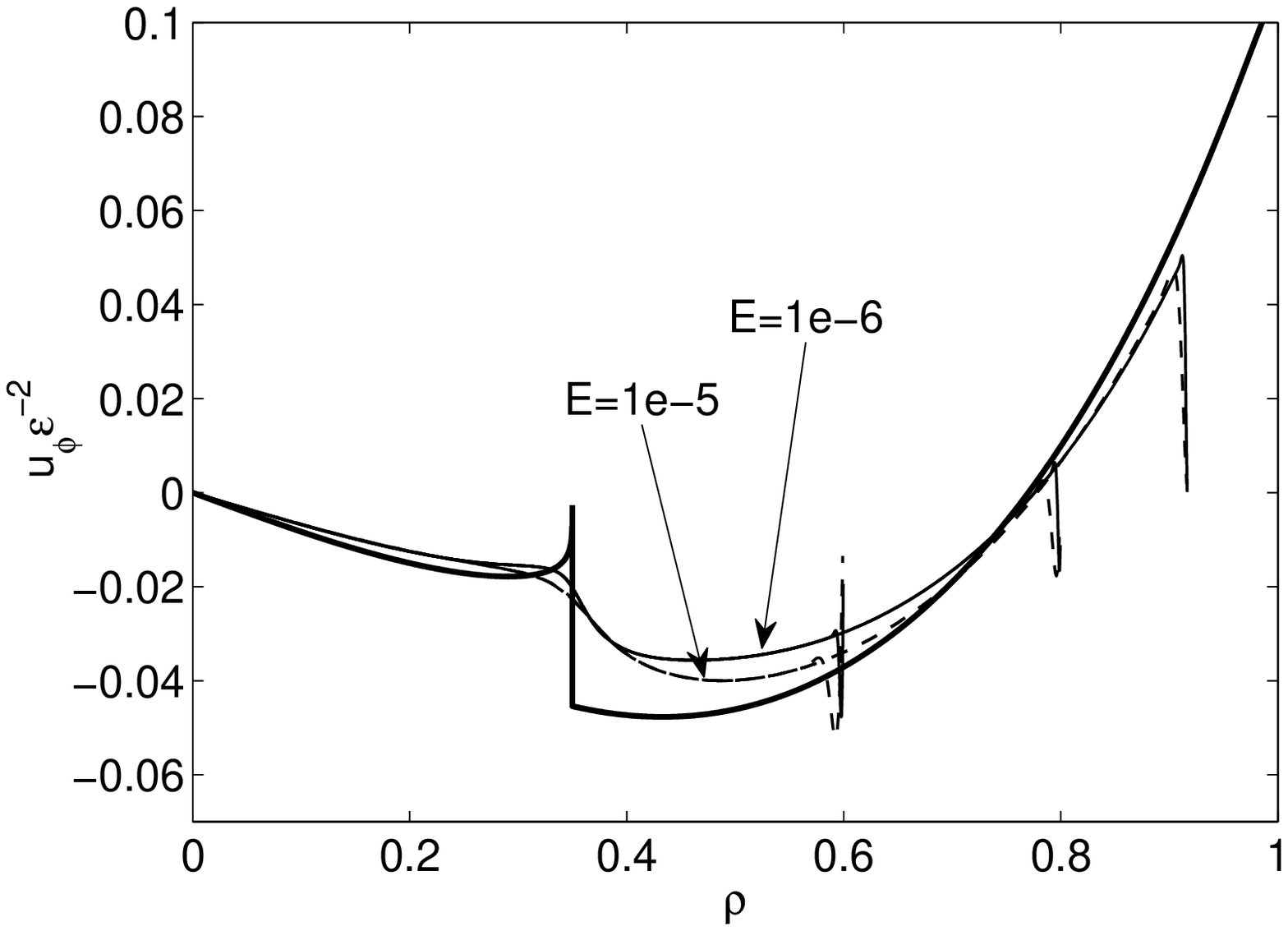} \\[0.5cm] (b) \hspace{11cm} ~\\[-1cm]
\includegraphics[width=10cm]{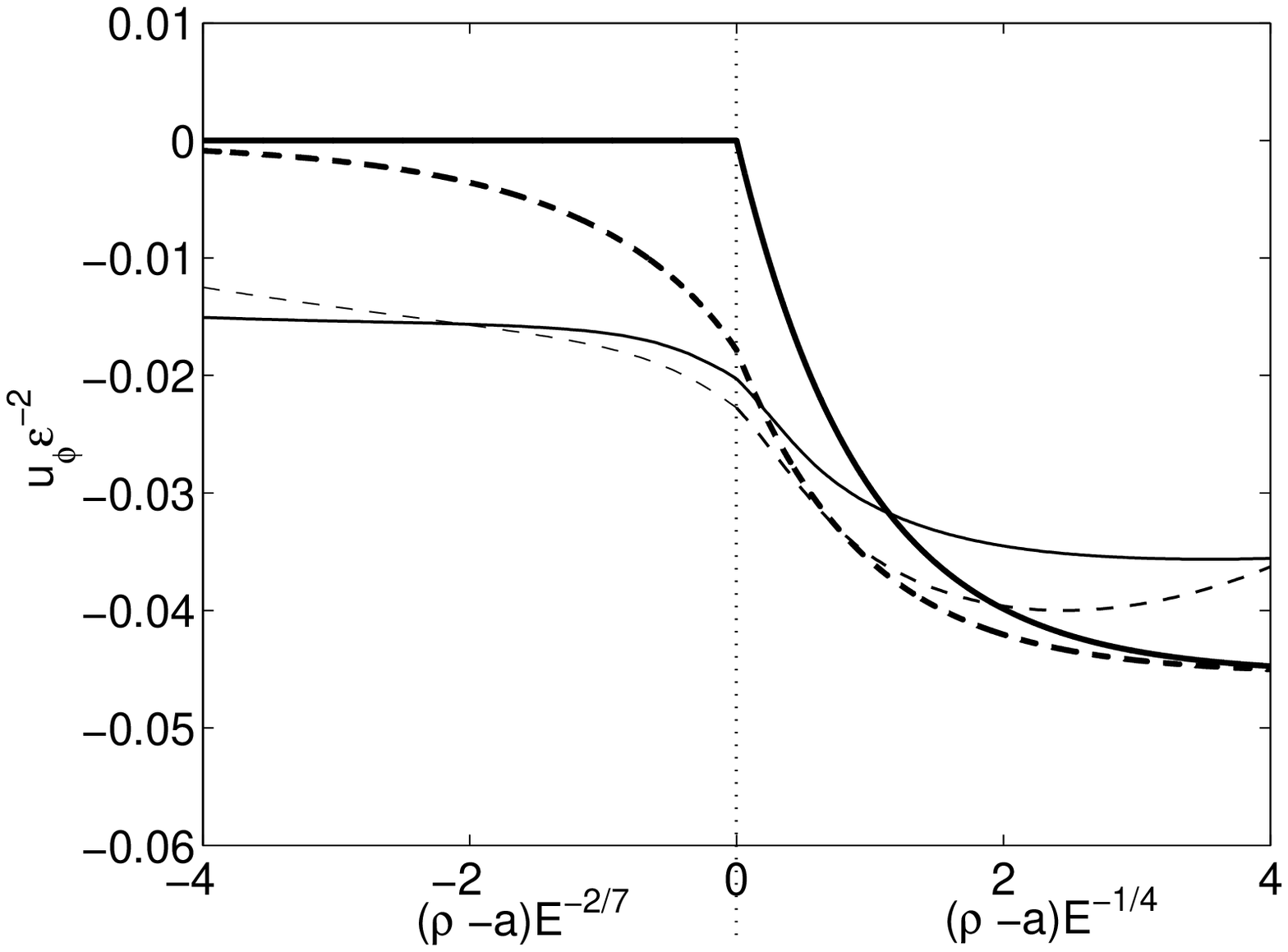} 
\end{center}
\caption{Mean azimuthal velocity  for  $\eps= 2.2 ~10^{-3}$, $a=0.35$, $\om=2.2$ and $\alpha=0$ (no libration ot the inner core). (a) Solution in the bulk: $u_{\phi_2}=u_{\phi}\eps ^{-2}$ versus $\rho$. The thick solid line is the theoretical prediction in the bulk (regions I and II). The thin lines are the numerical results obtained by \cite{calkins2010} at three different $z$ ($z=0.4, 0.6 , 0.8$) and two different Ekman numbers ($E=10^{-5}$: dashed lines, $E= 10^{-6}$: solid lines). 
(b)  Solution in the Stewartson layers:  $u_{\phi_2}=u_{\phi}\eps ^{-2}$ versus  $(\rho -a)/E^{1/4}$ for $\rho>a$ and versus $(\rho-a)/E^{2/7}$ for $\rho<a$. The thick solid line is the leading order theoretical prediction in the outer Stewartson layers. The thick dashed line is the theoretical prediction for $E=10^{-6}$ where the correction in $E^{1/28}$ has been taken into account. The thin lines are the numerical results of \cite{calkins2010} at $z=0.6$; solid line: $E=10^{-6}$, dashed line: $E=10^{-5}$.}
\label{fig:comparuphi}
\end{figure}
\end{center}
In a librating shell, we expect from the theory that the azimuthal zonal flow remains independent of $z$ in the bulk  but  exhibits a 
discontinuous behavior across the  cylinder tangent to the inner sphere at $\rho=a$. Such a discontinuity is visible in the numerical
results. In figure \ref{fig:comparuphi}(a), the numerical results obtained by \cite{calkins2010} for two values of the Ekman number 
($E= 10^{-5}$ and $10^{-6}$)  are compared with the theoretical prediction in the bulk  (regions I and II).  
 \begin{center}
\begin{figure}
\begin{center}
(a) \hspace{11cm} ~\\[-1cm]
\includegraphics[width=10cm]{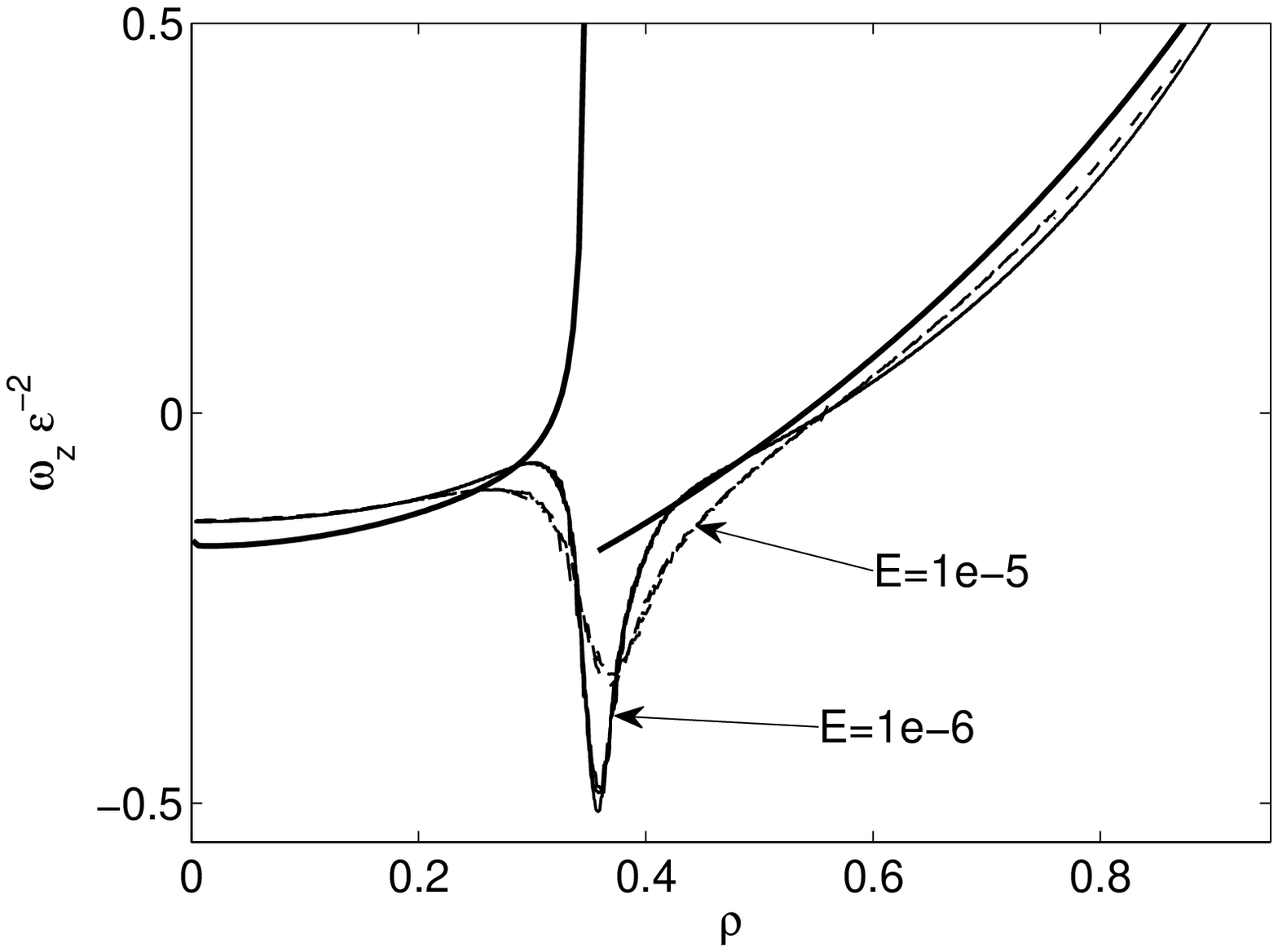} \\[0.5cm] (b) \hspace{11cm} ~\\[-1cm]
\includegraphics[width=10cm]{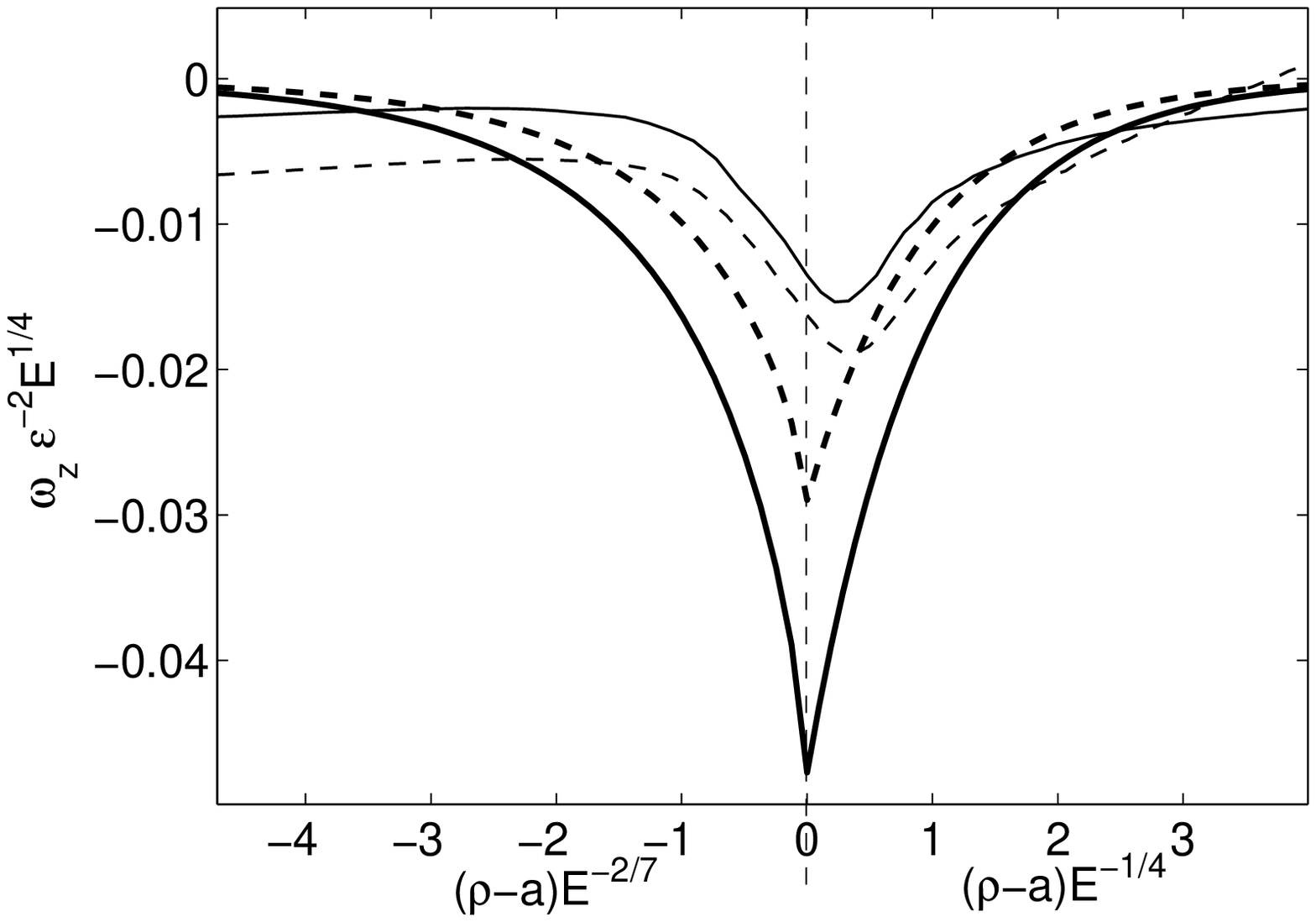}
\end{center}
\caption{Mean axial vorticity for $\eps= 2.2 ~10^{-3}$, $a=0.35$, $\om=2.2$ and $\alpha=0$ (no libration of the inner core).  (a) Solution in the bulk: $\om_{z}\,\eps ^{-2}$ versus $\rho$. The thick solid line is the theoretical prediction in the bulk (regions I and II). The thin lines are the numerical results obtained by \cite{calkins2010} at three different $z$ ($z=0.4, 0.6 , 0.8$) and two different Ekman numbers ($E=10^{-5}$: dashed lines, $E=10^{-6}$: solid lines). 
(b)  Solution in the Stewartson layers:  $\om_{z}\eps ^{-2}E^{1/4}$ versus  $(\rho -a)/E^{1/4}$ for $\rho>a$ and versus $(\rho-a)/E^{2/7}$ for $\rho<a$.  The thick solid line is the leading order theoretical prediction in the outer Stewartson layers. The thick dashed line is the theoretical prediction for $E=10^{-6}$ where the correction in $E^{1/28}$ has been taken into account. The thin lines are the numerical results of \cite{calkins2010} at $z=0.6$; solid line: $E=10^{-6}$, dashed line: $E=10^{-5}$.}  
\label{fig:comparwz}
\end{figure}
\end{center}
The agreement is good if we do not consider the region close to $\rho=a=0.35$ where the theoretical solution in the bulk is discontinuous.  
In particular, note that for both Ekman numbers, and any value of $z$, the azimuthal velocity profiles collide far from $\rho=0.35$ on a single curve, as expected from the theory. 
As in figure \ref{fig:syst_sphere}, the fast localized variations observed on the numerical curves correspond to the boundary layer on the outer sphere $\rho^2 + z^2 =1$. 
Their position changes  because different $z$ are considered. 
In the Stewartson layers, the solution becomes dependent of the Ekman number.  In figure \ref{fig:comparuphi}(b)
the numerical results for $z=0.6$   are plotted with respect to  the outer layer variables $(\rho -a)/E^{1/4}$ and $(\rho-a)/E^{2/7}$.
These results are compared to the leading order theoretical prediction as well as  to the two-order approximations obtained for $E=10^{-6}$ in this plot.
We can first note that there is a noticeable effect of the second order correction on the theoretical curve. This correction which is of order $E^{1/28}$ is clearly not small
for an Ekman number $E=10^{-6}$.  If this correction is taken into account, a relatively good agreement is obtained between the 
theoretical curve and the numerical results close to $\rho =a$. 

\begin{center}
\begin{figure}
\begin{center}
(a) \hspace{11cm} ~\\[-1cm]
\includegraphics[width=10cm]{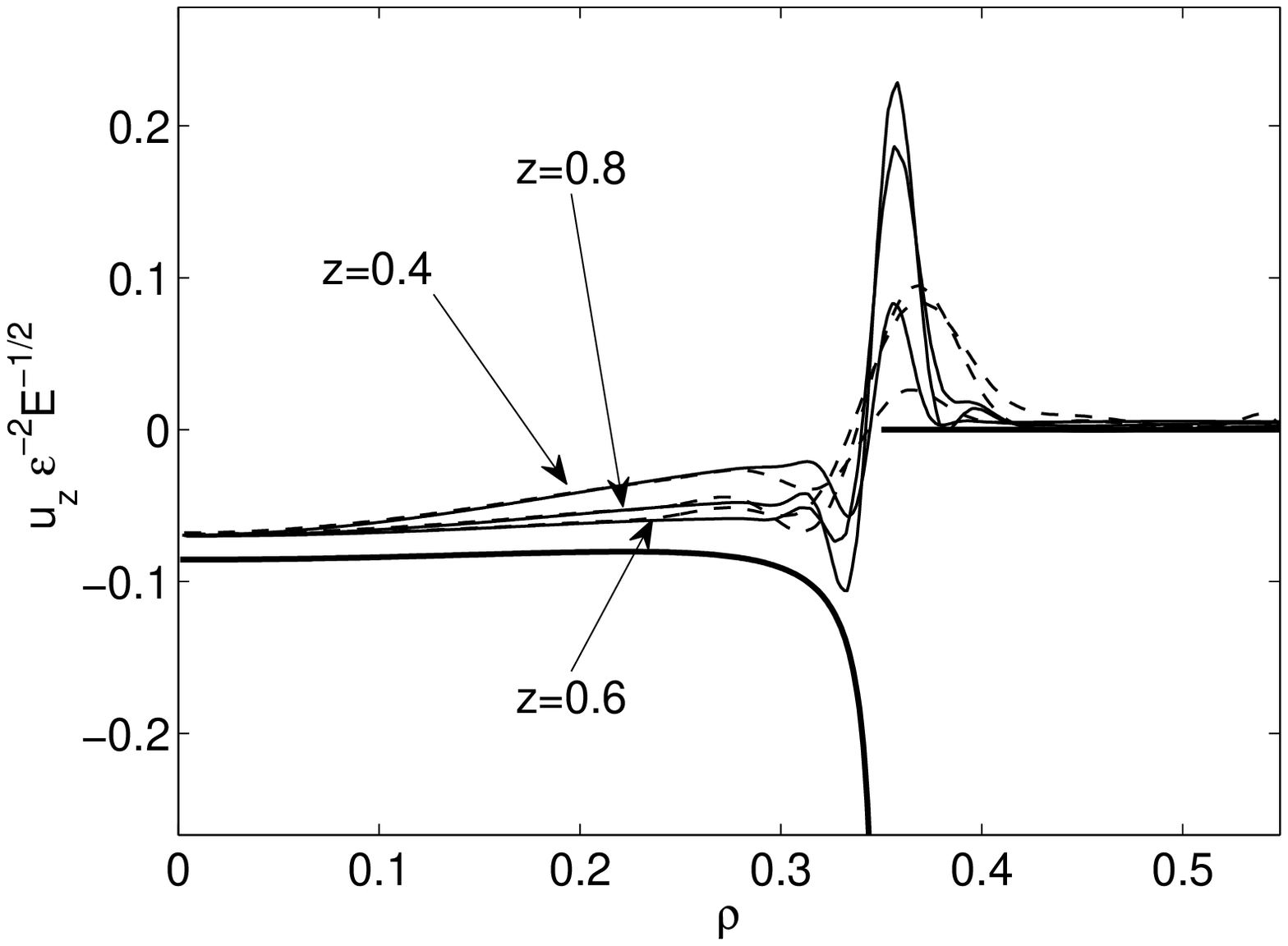} \\[0.5cm] (b) \hspace{11cm} ~\\[-1cm]
\includegraphics[width=10cm]{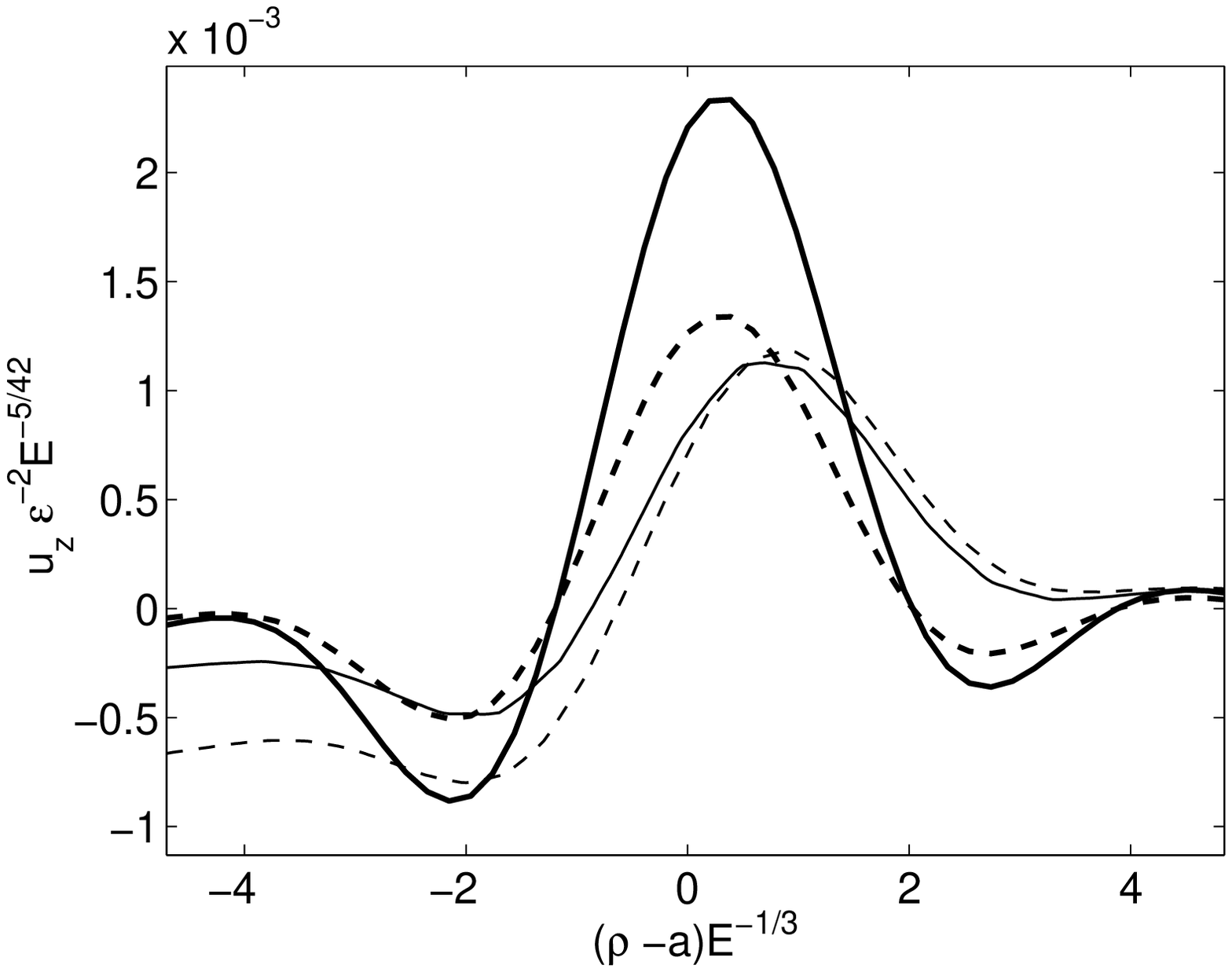}
\end{center}
\caption{Mean axial velocity  for $\eps= 2.2 ~10^{-3}$, $a=0.35$, $\om=2.2$ and $\alpha=0$ (no libration of the inner core). (a) Solution in the bulk: $u_{z} \,\eps ^{-2}E^{-1/2}$ versus $\rho$.
 The thick solid line is the theoretical prediction in the bulk (regions I and II). The thin lines are the numerical results obtained by \cite{calkins2010} at three different $z$ ($z=0.4, 0.6, 0.8$) and two different Ekman numbers ($E=10^{-5}$: dashed lines, $E=10^{-6}$: solid lines). 
   (b) Solution in the inner Stewartson layer at $z=0.7$:  $u _{z}\,\eps ^{-2}E^{-5/42}$ versus  $(\rho -a)/E^{1/3}$. The thick solid line is the leading order theoretical prediction in the inner Stewartson layer. The thick dashed line is the theoretical prediction for $E=10^{-6}$ where the correction in $E^{1/28}$ has been taken into account. The thin lines are the numerical results of \cite{calkins2010} at $z=0.7$; solid line: $E=10^{-6}$, dashed line: $E=10^{-5}$. }
\label{fig:comparuz}
\end{figure}
\end{center}
A similar comparison is performed for the axial vorticity  in figure \ref{fig:comparwz}   and for the axial velocity in figure \ref{fig:comparuz}. 
In figures \ref{fig:comparwz}(a)  and \ref{fig:comparuz}(a), we have compared the numerical results with the theoretical solution in the bulk, that is (\ref{exp:uphi2Bulk}) with
(\ref{exp:Psi2II}a,b) in region II ($r<a$), and with (\ref{exp:Psi2I}), (\ref{exp:Bchi2I}) in region I ($r>a$).  A good agreement is observed far from $\rho=a$. 
Note in particular that both axial vorticity and axial velocity are independent of $z$ far from $a$ as expected from the theory.
However, there are systematic differences which were
not present in the sphere. We suspect that these differences are due to $O(E^{1/28})$ corrections  generated in the Stewartson layers, and  
which are not negligible for $E=10^{-6}$. 
In figure \ref{fig:comparwz}(b), the axial vorticity is plotted with respect to the  outer-layer variables.  Although a slight shift is observed in the numerical results, we 
can see that the vorticity peak with its scaling in $E^{-1/4}$  is qualitatively described.  
In figure \ref{fig:comparuz}(b), the axial velocity is plotted with respect to the inner layer variable. The agreement of the numerical results with the second-order 
theoretical approximation is remarkable for this value of $z$. Again a slight shift is observed which seems to diminish as $E$ decreases.
Other values of $z$ have also been tested and a less good agreement has been observed for smaller values of $z$. The numerical results
shows a weaker jet whereas the theory predicts the opposite.  We suspect that this qualitative difference could be due not only to the 
large value of the Ekman number but also to the particular structure of the Stewartson layer for small $z$ which has not been 
resolved here.



\section{Conclusion}

In this paper, we have analyzed the mean flow induced by longitudinal libration (that is small oscillation of the rotation rate)  in a  spherical shell and in a sphere,  when the oscillation frequency
is larger than twice the mean rotation rate. 
Using asymptotic methods in the limit of small Ekman numbers $E$ and small oscillation amplitude $\eps$, we have been able to show that the 
main contribution to the mean flow is,  in the bulk, an azimuthal flow which scales as $\eps^2$, remains independent of the Ekman number and depends on the cylindrical variable $\rho$ only. We have shown 
how such a zonal flow is generated by nonlinear interaction of the harmonic solution in the boundary layers. 
The dependence of the zonal flow with respect to the libration frequency $\om$ and the ratio of the core size $a$ has also  been comprehensively analysed.
In particular, we have  shown how the zonal flow in a shell can be deduced from the zonal flow in a sphere.  
We have also seen that for moderate frequencies $2<\om<8.5$, both cyclonic and anticyclonic mean rotation can be created by libration,
whereas only weak cyclonic rotation is created for $\om >8.5$.  

The presence of the inner sphere has also been shown to create a discontinuity of the azimuthal flow across the tangent cylinder at $\rho=a$ 
which is smoothed in a series of nested viscous layers of widths $E^{1/4}$, $E^{2/7}$ and $E^{1/3}$, as for the configuration of differentially rotating spheres
\cite[][]{stewartson1966}.  The mean flow correction has been calculated in each of these layers. 
We have obtained that an axial vorticity field of order $\eps^2 E^{-1/4}$ is created in the outer Stewartson layers,
and that a non-uniform axial flow of order $\eps^2 E^{5/42}$ is  present in the inner layer ($|\rho -a| =O(E^{1/3})$). 
We have shown that the spatial structures of these two fields in the Stewartson layers are universal functions which do not depend (after rescaling) 
on any parameter. The ratio of the core size, the oscillation frequency as well as the relative oscillation amplitude of the 
inner sphere, have been shown only to intervene  in an amplitude factor which has been fully characterized. In particular,  
we have demonstrated that the libration of the inner sphere adds a very simple quantity to the amplitude factor obtained without inner sphere libration, which
always  tends to increase the anticyclonicity of the vorticity field, and  to add an axial flow from the inner core to the outer sphere.

The asymptotic results have been compared to numerical results of the literature \cite[][]{calkins2010} or obtained with a finite-element code,
and a good agreement has been demonstrated for the zonal flow in  the bulk. A reasonable agreement has also been obtained for the peculiar structure
of the solution in the Stewartson layers despite the largeness of the Ekman number in the simulations. 
In particular, we have shown that it was necessary to consider the second-order $O(E^{1/28})$ correction to obtain a correct agreement.  
It is worth mentioning that  our theory is valid up to $O(E^{1/14})$. Even for realistic astrophysical values of $E$ ($E\approx 10^{-14}$), 
we cannot guarantee that the error will remain small. This constitutes a weak point of the theory that is important to keep in mind when
the results are applied.

\section{Discussion}

The present paper has considered oscillation frequencies larger than twice the rotation rate in a spherical shell. 
We can naturally address the question of the libration response when 
 the frequency is smaller than twice the rotation rate
or when the boundaries of the shell are elliptically deformed (that is the outer/inner spheres are ellipsoids). 

The first point has already been addressed in a cylinder \cite[][]{wang1970,sauret2012} and for vanishingly small frequencies
in a sphere  \cite[][]{busse2010,sauret2010}.  For this point, the  first difficulty comes from the fact that 
inertial modes can be resonantly excited by the harmonic forcing \cite[][]{morize2010,tilgner2007}.  When this occurs, the 
zonal flow is strongly modified and is no longer dominated by the nonlinear interaction in the boundary layers \cite[][]{sauret2012}.
The second difficulty is the singular behavior of the boundary layer solution when twice the cosine of the inclination angle of the boundary
matches  the oscillation frequency. This singularity is known to generate internal shear layers in the bulk \cite[see][]{kerswell1995,kida2011}
whose effects on the mean flow remain to be quantified.

\begin{figure}
\begin{center}
\includegraphics[width=8cm]{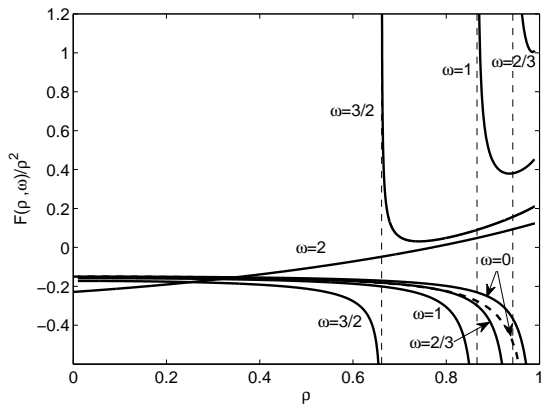} 
\end{center}
\caption{The function $\BF(\rho;\om)/\rho^2$ versus $\rho$ for different values of $\om$ between $0$ and $2$ . The function $\BF$ diverges at $\rho_c= \sqrt{1-\om^2/4 }$. 
Expressions (\ref{exp:BF+}) and (\ref{exp:BF-}) have been used for $\BF$ for $\om >0$. The dashed line for $\om =0$ is the expression (\ref{exp:FBusse}) obtained by \cite{busse2010}.}
\label{fig:Fwpetit}
\end{figure}
If we neglect the effect of such internal shear layers and assume no resonance, the mean zonal flow can be calculated using the same approach as for $\om>2$. The 
expressions (\ref{exp:Psi2I}, \ref{exp:Bchi2I}) and (\ref{exp:Psi2II}a,b) for $\Psi_2$ and $\Bchi _2$ still apply but  expression (\ref{exp:BF+}) for $\BF(\rho;\om)$ obtained for $\om >2$ 
can only be used for $\rho > \rho_c= \sqrt{1-\om^2/4}$. 
For $\rho < \rho_c$,  expression (\ref{exp:BF-}) has to be used. This expression is obtained by the same analysis by changing the definition of the square root 
in (\ref{exp:lambda}). 
Both expressions diverges at $\rho_c$ which corresponds to the radius of the critical latitude on the outer sphere. 
In a shell, this singularity in $\BF$ implies two singularities in $\Bchi_2$. The second one which is associated with the critical latitude on
the inner sphere is at $\rho_c= a \sqrt{1-\om^2/4}$.
From an asymptotic point of view, these singularities mean that another scaling has to be found for $\Bchi_2$ close to each singularity.  

In figure \ref{fig:Fwpetit}, we have plotted $\BF /\rho^2$, that is the angular velocity of the zonal flow in region I (see figure \ref{fig:syst_noyau}), 
for different values of $\om$. 
It is interesting to note that for frequencies ranging from $0$ to $1$, we 
obtain almost the same constant retrograde rotation for all $\rho < 0.6$. 

As $\om \rightarrow 0$, expression (\ref{exp:BF-})  reduces   to 
\be
\BF(\rho;\om) \sim \frac{(59\,\rho^2-72)\rho^2}{480\,(1-\rho^2)}.
\label{exp:F0} 
\ee
This expression slightly differs from the expression obtained by  \cite{busse2010} which reads
\be
\BF(\rho;\om) \sim \frac{(51.8 \,\rho^2-72)\rho^2}{480\,(1-\rho^2)} .
\label{exp:FBusse} 
\ee
 Expression (\ref{exp:F0}) is valid when $\eps \ll \om \ll 1$ whereas Busse expression  applies when
 $\om \ll \eps$.  As noticed in figure \ref{fig:Fwpetit}, the difference between these two expressions is 
particurlarly visible for $\rho>0.7$. 

The small frequency limit is interesting because 
small frequency inertial modes are generally too  damped  to be resonantly excited. 
Moreover, for small $\om$, internal shear layers are limited to the very small region near the equator and
to the neighborhood of the cylinder tangent to the inner sphere,  that is within
  the Stewartson layers. Therefore, they are disconnected from the bulk regions I and II 
  where our solution is expected to apply.

The second point, that is the libration response in an ellipsoid, has also been addressed in the literature \cite[][]{Chan2011,Zhang2011}. 
When the ellipsoid is axisymmetric with respect to the rotation axis, our asymptotic analysis can be 
very easily adapted to find the zonal flow which will not be fundamentally different from the one in the sphere.
By contrast, when the container is no longer axisymmetric along the rotation axis, the driving mechanism 
of the zonal flow may change as it may no longer be associated with nonlinear interaction of  boundary layer solution but to
nonlinear interaction in the bulk of pressure forced solutions.

The two issues considered in this section are also discussed in the recent paper by \cite{Noir2012}.

We would like to thank  M.~A.~Calkins for having provided the numerical data published in \cite{calkins2010} for the comparison done in section 5.  
This work has  benefitted from discussions with M.~Le Bars,  D.~C\'ebron, and F.~Busse.  
We would also like to acknowledge  discussions with N. Rambaux and C. Moutoux concerning the astrophysical applications of our results 
to Galilean satellites and exoplanet systems. 


\newpage

\appendix

\section{The nonlinear coefficients $A_l$, $B_l$ , $C_l$  and $D_l$}

The coefficients appearing in expressions  (\ref{exp:NL}a,b) 
are given by 
 \begin{subeqnarray}\label{coeffAB}
 A_{1}  &= & \frac{\sin ^2\theta}{16\, {s_-}^3 \,{{s_+}}^3}\left(4 {s_-}^2 \,({s_-} + \text{i}\, {s_+})^2 \,{s_+}^2 \,\cos^2\theta + (-{s_-}^6 + {s_-}^4\, {s_+}^2 - 4  \text{i}\, {s_-}^3\, {s_+}^3 - {s_-}^2\, {s_+}^4 + 
      {s_+}^6)\, \sin ^2\theta  \right. \nonumber  \\ 
 & &  \qquad \qquad \qquad \qquad 
\left. + \cos\theta (-4 {s_-}^2 {s_+}^2 ({s_-}^2 + {s_+}^2)^2 + ({s_-}^4 + {s_+}^4)\sin ^2\theta )\right) ,\\
 B_{1}  & =   &
-\frac{(1- \text{i})}{16 {s_-}^2 {s_+}^2} ({s_-}^2 +  {s_+}^2) ( {s_-} + \text{i} {s_+}) (({s_-} - \text{i} {s_+})^2 - \cos\theta])\sin ^4\theta ,
\\
 A_{2}  & = & \frac{\sin^2 \theta\cos\theta}{16\,{s_+}^2}\left(4\,{s_+}^2\,\cos\theta+\sin^2\theta\right) ,\\
  A_{3}  & = & \frac{\sin^2 \theta\cos\theta}{16\,{s_-}^2}\left(-4\,{s_-}^2\,\cos\theta+\sin^2\theta\right) , \\
A_{4} &  = & - \frac{\sin^2\theta }{16{s_-}^3{s_+}^2}({s_+}^2+\cos\theta)\left(4\cos\theta\,({s_-}-{s_+}){s_+}^2{s_-}^2 +\sin^2\theta ({s_+}^3+{s_-}^3)\right) ,\\
 A_{5} & = & \frac{\sin^2\theta }{16{s_-}^2{s_+}^3}({s_-}^2-\cos\theta)\left(4\cos\theta\,({s_-}-{s_+}){s_+}^2{s_-}^2 +\sin^2\theta ({s_+}^3+{s_-}^3)\right), \\
  B_{2} & =& B_{3} = B_{4} =B_{5}  =  0 ,\\
C_{1} & = & 
\frac{(1- \text{i})\sin^2\theta}{32 {s_-}^3 {s_+}^3}
  \left(({s_-}^5 - {s_-}^3 {s_+}^2 - \text{i} {s_-}^2 {s_+}^3 + \text{i} {s_+}^5) \sin^2\theta  \right. \nonumber  \\ 
      && \qquad \qquad \qquad \qquad   \left.+ 4 {s_-}^2 ({s_-} - \text{i} {s_+}) ({s_-}^2 + {s_+}^2) {s_+}^2 \cos\theta\right) ,\\
 D_{1} & = &\frac{\text{i}\,\sin^4\theta}{16 {s_-}^2{s_+}^2}({s_+}^2+{s_-}^2)^2, \\
 C_{4}  &=& \frac{(1- \text{i})\sin^2\theta }{32{s_-}^3{s_+}}\left(4\cos\theta\,({s_-}-{s_+}){s_+}^2{s_-}^2 +\sin^2\theta ({s_+}^3+{s_-}^3)\right) ,\\
   C_{5}  &=& -\frac{(1- \text{i})\sin^2\theta }{32{s_+}^3{s_-}}\left(4\cos\theta\,({s_-}-{s_+}){s_+}^2{s_-}^2 +\sin^2\theta ({s_+}^3+{s_-}^3)\right) ,\\
 C_{2}& = & D_{2} = 
 C_{3} = D_{3}= D_{4}=D_{5} = 0 .
 \end{subeqnarray}
 \noindent where 
  \be
  {s_+} = \sqrt{\frac{\omega}{2}+\cos\theta}, \qquad {s_-} = \sqrt{\frac{\omega}{2}-\cos\theta}.
 \ee

%

\section{The function $\BF(\rho;\om)$}

For $\om >2$, we have the following expression for $\BF(\rho;\om)$ with $\rho=\sin\theta$:
\begin{eqnarray}
\BF (\sin\theta;\om) & = & \Biggl[\frac{({s_+}^3+{s_-}^3)\sin^4\theta}{32\,{s_+}^2\,{s_-}^2}+\frac{({s_-}-{s_+})\,\cos\theta \sin^2\theta}{8} \Biggr] 
 \,\Biggl[\frac{{s_+}-\sqrt{\cos\theta}}{{s_+}^2+\cos\theta}-\frac{{s_-}-\sqrt{\cos\theta}}{{s_-}^2-\cos\theta}\Biggr] \nonumber \\
& & +\frac{\sin^2\theta ({s_+}+\sqrt{\cos\theta})}{32\,{s_+}^3(\cos^2\theta+4\,{s_+}^4)}[2\,{s_+}^2-2\,{s_+}\,\sqrt{\cos\theta}+\cos\theta]\,[\sin^2\theta+4\,{s_+}^2\,\cos\theta] \nonumber \\
 & & +\frac{\sin^2\theta ({s_-}+\sqrt{\cos\theta})}{32\,{s_-}^3(\cos^2\theta+4\,{s_-}^4)}[2\,{s_-}^2-2\,{s_-}\,\sqrt{\cos\theta}+\cos\theta]\,[\sin^2\theta-4\,{s_-}^2\,\cos\theta] \nonumber \\
 && -\frac{\sin^2\theta}{32\,{s_-}^3\,{s_+}^3\,({s_-}^2+{s_+}^2)\,[{s_-}^2+{s_+}^2+2\,{s_+}\,\sqrt{\cos\theta}+\cos\theta]^2} \nonumber \\ 
&& \Bigl[ ({s_-}^2+{s_+}^2)[{s_-}^6+7\,{s_+}^2\,{s_-}^4+8\,{s_-}^7\,{s_+}^3-7\,{s_-}^2\,{s_+}^4+16\,{s_-}^5\,{s_+}^5\nonumber \\
&& -{s_+}^6+4\,{s_+}^3\,{s_-}^3\,(-3+2\,{s_+}^4)]  +[4 {s_-}^6 {s_+}-{s_-}^4 {s_+}^3-16 {s_-}^2 {s_+}^5-3 {s_+}^7\nonumber \\
&&+4 {s_-}^5 {s_+}^2 (-3+16 {s_+}^4)+{s_+}^4 {s_-}^3 (-19+32 {s_+}^4)  +{s_-}^7 (-1+32 {s_+}^4)] \sqrt{\cos\theta} \nonumber \\
 & &
+[4 {s_-}^8 {s_+}^2-3 {s_+}^6+{s_-}^6 (1-20 {s_+}^4)+{s_-}^2 {s_+}^4 (-11+4 {s_+}^4)\nonumber \\
&& +4 {s_+}^3 {s_-}^3 (-3+16 {s_+}^4)+{s_-}^5 (-4 {s_+}+64 {s_+}^5)+{s_-}^4 ({s_+}^2-20 {s_+}^6)] \cos\theta \nonumber \\
&& -[4\,{s_-}^7\,{s_+}^2+4\,{s_-}^6\,{s_+}^3+{s_+}^5+56\,{s_-}^4\,{s_+}^5+{s_-}^3\,{s_+}^2\,(3-76\,{s_+}^4) \nonumber \\
& & +{s_-}^5\,(1-8\,{s_+}^4)
+3\,{s_+}^3\,{s_-}^2\,(1-4\,{s_+}^4)]\,\cos^{3/2}\theta \nonumber \\
& & 
+[-{s_-}^8-4 {s_-}^6 {s_+}^2-4 {s_+}^3 {s_-}^5-48 {s_-}^4 {s_+}^4 +60 {s_+}^5 {s_-}^3+20 {s_-}^2 {s_+}^6+{s_+}^8] \cos^2\theta \nonumber \\
&&
+[{s_-}^7-4 {s_-}^6 {s_+}+8 {s_-}^5 {s_+}^2-11 {s_-}^4 {s_+}^3+31 {s_+}^4 {s_-}^3  +20 {s_-}^2 {s_+}^5+3 {s_+}^7] \cos^{5/2} \nonumber\\
&&
+[-{s_-}^6+4\,{s_-}^5\,{s_+}-{s_+}^2\,{s_-}^4+12\,{s_+}^3\,{s_-}^3+11\,{s_-}^2\,{s_+}^4+3\,{s_+}^6]\,\cos^3\theta \nonumber\\
&&
+ ({s_-}^5+3\,{s_-}^3\,{s_+}^2+3\,{s_+}^3\,{s_-}^2+{s_+}^5)\,\cos^{7/2}\theta 
\Bigr] \nonumber \\
& & +\frac{\sin^2\theta}{4} ,
\label{exp:BF+}
\end{eqnarray}
\noindent where
\begin{equation}
{s_+}=\sqrt{\frac{\omega}{2}+\cos\theta} \quad \text{and} \quad {s_-}=\sqrt{\frac{\omega}{2}-\cos\theta} .
\end{equation}
For $\om<2$, the  above expression  applies  when $\rho > \rho_c=\sqrt{1-\om^2/4}$.
When $\rho < \rho_c$, we have to use the following expression with $\rho=\sin\theta$
\begin{eqnarray}
\BF (\sin\theta;\om) & = &\frac{\sin^2\theta}{32}\Biggl[8+ \frac{(l_++\sqrt{\cos\theta})\,(4\,l_+^2\,\cos\theta+\sin^2\theta)}{l_+^3\,(2\,l_+^2+2l_+\,\sqrt{\cos\theta}+\cos\theta)}+ \frac{(l_-+\sqrt{\cos\theta})\,(4\,l_-^2\,\cos\theta+\sin^2\theta)}{l_-^3\,(2\,l_-^2+2l_-\,\sqrt{\cos\theta}+\cos\theta)} \nonumber\\
& & -\frac{4\,(l_-^3+l_-^2\,\sqrt{\cos \theta}+(l_-+l_+)\,\cos\theta}{l_+\,l_-\,(l_-+\sqrt{\cos\theta})}+\frac{l_-^4\sin^2\theta\,(l_-+\sqrt{\cos\theta}+(l_+^3+l_-^3)/l_-^4\,\cos \theta)}{\cos\theta\,l_+^3\,l_-^3\,(l_-+\sqrt{\cos\theta})} \nonumber \\
& & -\frac{4\,(l_+^3+l_+^2\,\sqrt{\cos \theta}+(l_++l_-)\,\cos\theta}{l_-\,l_+\,(l_++\sqrt{\cos\theta})}+\frac{l_+^4\sin^2\theta\,(l_++\sqrt{\cos\theta}+(l_-^3+l_+^3)/l_+^4\,\cos \theta)}{\cos\theta\,l_-^3\,l_+^3\,(l_++\sqrt{\cos\theta})} \nonumber \\
&&+ \frac{1}{\cos\theta\,l_-^3 \,l_+^3 \,(l_- + l_+) \,(l_- + l_+ + \sqrt{\cos\theta})^2} \nonumber \\
&& \Biggl(4 l_-^2\,l_+^2 \, (l_- - l_+)^2\, \cos^{5/2}\theta\,[1+2\,(l_- + l_+)\, \sqrt{\cos\theta}\,] \nonumber \\
&&+  \cos\theta \,[l_- \,l_+ \,(l_- + l_+)\, (2\,(2 l_-^2 - 3 l_- l_+ +  2 l_+^2)\,\sin^2\theta+ 4 l_-^5 l_+  - 8 l_-^3 l_+^3 + 4 l_- l_+^5 )\nonumber \\
&& +  2 \,(l_-^5 + l_+^5) \sin^2\theta] + (l_- + l_+)^2\,(l_-^4 + l_+^4) \,\sin^2\theta \,[ (l_- + l_+)+2\, \sqrt{\cos\theta}]\nonumber \\
&&+ \cos^{3/2}\theta\,[\,8 \,l_+^2\,l_-^2\, (l_-^2 - l_+^2)^2 +(l_-^4 + l_+^4 + l_- l_+ \,(l_-- l_+)^2) \sin^2\theta] \Biggr)\Biggr]
\label{exp:BF-}
\end{eqnarray}
\noindent where
\begin{equation}
{l_+}=\sqrt{\cos\theta+\frac{ \omega}{2}} \quad \text{and} \quad {l_-}=\sqrt{\cos\theta-\frac{\omega}{2}} .
\end{equation}

\section{Stewartson layers}

In this section, the solutions in the different layers around $\rho=a$ are provided. The analysis mainly follows the derivation given by \cite{stewartson1966}.
In all the layers, the functions $\psi_2$ and $\chi_2$ are related at leading order in $E$ by a same set of equations which can be written in 
cylindrical coordinates as
\bsea 
2\,\frac{\partial \chi_2}{\partial z}=E\,\frac{\partial^4\psi_2}{\partial \rho^4}   ,\\
-2\,\frac{\partial \psi_2}{\partial z}=E\,\frac{\partial^2\chi_2}{\partial \rho^2} .
\label{motion11}
\esea

In the two outer layers (see figure  \ref{fig:Stewartson}), the function $\chi _2$ remains at leading order in $E$ independent of $z$, such that 
a simple relation between $\psi_2$ and $\chi _2$ can be obtained by integrating (\ref{motion11}b) with respect to $z$:
 \begin{equation}
\psi_2(\rho,z)=-E\,\frac{z}{2}\,\frac{\partial^2 \chi_2}{\partial \rho^2}+ B(\rho) ,
\label{stewartson1}
\end{equation}
where $B(\rho)$ is a function to be determined  in each outer layer.

\subsection*{External outer layer $E^{1/4}$}

In the external outer layer of thickness $E^{1/4}$,  the function $\psi_2$ must be antisymmetric with respect to the 
equator $z=0$. This implies $B(\rho)=0$ in this layer.   
Another relation is obtained by considering the problem close to the boundary layer at $z=\sqrt{1-a^2}$. In this area,
a relation between $\psi_2$ and $\chi_2$ is deduced by writing (\ref{relation(o)}) close to $\sin\theta= a$
\be
\chi_2(\rho) = -2 \,(1-a^2)^{1/4}\, E^{-1/2}\,\psi_2(\rho,\sqrt{1-a^2}) + \BF(a;\om) .
\label{exp:chi2ExOut}
\ee
Evaluating (\ref{stewartson1}) at $z=\sqrt{1-a^2}$ then leads  to the equation  
\be
\frac{\partial^2 \chi_2}{\partial \rho^2}-\frac{\chi_2-\BF(a;\om)}{{E}^{1/2}\,(1-a^2)^{3/4}}=0,
\ee
which immediatly gives
\begin{eqnarray}
\chi_2(\rho) = \BF(a;\om)-\Lambda_1\,\text{exp}\Biggl[-\frac{(\rho-a)\,E^{-1/4}}{(1-a^2)^{3/8}}\,\Biggr].
\end{eqnarray}
At this level, the constant $\Lambda_1$ is, a priori, unknown. 
We shall see below that it is given at leading order in $E$ by 
\be
\Lambda_1 = \delta \Bchi = \Bchi^+ - \Bchi ^- = \BF(a;\om)- \frac{\alpha ^2 a^2}{4 \om^2} .
\label{exp:A1}
\ee
In this layer, the axial vorticity is given by 
\be 
\omega_{z_2} = \frac{1}{\rho}\pa{\chi_2}{\rho}= E^{-1/4}\frac{\delta \Bchi(a,\om)}{a\,(1-a^2)^{3/8}}\,\text{exp}\Biggl[-\frac{(\rho-a)\,E^{-1/4}}{(1-a^2)^{3/8}}\,\Biggr] ,
\label{exp:wzExOut}
\ee
while the function $\psi _2$ is given by the relation (\ref{stewartson1}) with $B(\rho)=0$:
\be 
\psi _2(\rho,z) = E^{1/2}\, \frac{z}{2}\,\frac{\delta\Bchi(a,\om)}{(1-a^2)^{3/4}} \,\text{exp}\Biggl[-\frac{(\rho-a)}{(1-a^2)^{3/8}\,E^{1/4}} \,\Biggr].
\label{exp:psi2ExOut}
\ee

\subsection*{Internal outer layer $E^{2/7}$}

The solution in the Internal outer layer is obtained by considering the problem close to the two boundaries (at $z=0$ and $z=\sqrt{1-a^2})$.
Close to the boundary on the inner sphere ($z=0$), we obtain a relation between $\psi_2$ and $\chi_2$ by using (\ref{relation(i)}):
\be
\chi_2(\rho) = 2 \left[1-(\rho/a)^2\right]^{1/4} E^{-1/2}\,\psi_2(\rho,0) + \frac{\alpha ^2\, a^2}{4\, \om ^2}  .
\label{exp:chi2InOut}
\ee
Considering (\ref{stewartson1}) at $z=0$, we then obtain  $B(\rho)$ in this layer:
\be 
B(\rho) = \frac{\chi_2(\rho) -  \alpha ^2\, a^2/(4 \om ^2)}{2\,[1-(\rho/a)^2]^{1/4}}\,\sqrt{E}.
\label{exp:B}
\ee

If we now use the relation (\ref{relation(o)}),   we obtain from  (\ref{stewartson1}), applied close to $z=\sqrt{1-a^2}$
with (\ref{exp:B})
\be
\frac{\partial^2 \chi_2}{\partial \rho^2}-\frac{\chi_2-\BF(a)}{(1-a^2)^{3/4}\,\sqrt{E}}-\frac{\chi_2-\alpha^2a^2/(4\,\om^2)}{[2(a-\rho)]^{1/4}}\frac{a^{1/4}}{\sqrt{E}\,\sqrt{1-a^2}} =0.
\label{exp:d2chi2}
\ee

This equation leads to the following solution  
\be
\chi_2(\rho) = \frac{\alpha^2 a^2}{4\,\om^2} +  E^{1/28}\Lambda_2\,\mathbf{f}(\brho) + O(E^{1/14}),
\label{exp:chi2}
\ee
with 
\be
\brho=(a-\rho)\Biggl[\frac{a}{2\,E^2(1-a^2)^2}\Biggr]^{1/7},
\label{exp:brho}
\ee
and 
\be
\Lambda_2=-\left(\frac{2}{a}\right)^{1/7}\frac{(1-a^2)^{-5/56}}{\mathbf{f}'(0)}\, \delta \Bchi(a,\om) , 
\label{exp:A2}
\ee
where the function $\mathbf{f}(s)$  satisfies
\be
\pdd{\bbf}{s} - \frac{\bbf}{s^{1/4}} =0 ,~~ ~\bbf(0)=1, ~~\bbf(\infty) =0 .
\label{equ:f}
\ee
An explicit expression for $\mathbf{f}(s)$ is provided by 
\be
{\bf f}(s)=\frac{2\,(4/7)^{11/7}\,s^{1/2}\,{\bf K}_{4/7}(8/7 s^{7/8})}{\Gamma(11/7)},
\label{exp:f}
\ee
where ${\bf K}_{\nu}$ is a modified Bessel function of order $\nu$ and $\Gamma(z)$ is  the Gamma function \cite[][]{Abramowitz_book}. 
Note that ${\mathbf{f}'(0)}=-(4/7)^{8/7}\,\Gamma(3/7)/\Gamma(11/7)\approx -1.2246$. 

The expressions (\ref{exp:A1}) and (\ref{exp:A2}) for $\Lambda_1$ and $\Lambda_2$ are such that 
$\chi _2$ and $\partial_\rho \chi_2$, obtained from (\ref{exp:chi2ExOut}) and (\ref{exp:chi2InOut}) for $\rho>a$ and $\rho<a$ respectively, 
are continuous at $\rho =a$.  
As shown by \cite{stewartson1966}, this continuity condition comes from the condition of matching with the solution in the inner layer. 
It guarantees the continuity of the axial vorticity which is given in the Internal outer layer by
\be
\om_{z_2} = E^{-1/4}\, \frac{\Lambda_2}{2^{1/7}\,a^{6/7}\,(1-a^2)^{2/7}} \,\bbf' (\brho) .
\label{exp:wzInOut}
\ee

Using (\ref{stewartson1}), (\ref{exp:B}), (\ref{exp:d2chi2}) and (\ref{exp:chi2}), an expression for $\psi_2$ can be obtained:
\be
\psi_2 \sim  E^{13/28} \left(1- \frac{z}{\sqrt{1-a^2}} \right)  \frac{\Lambda_2}{2^{9/7}\,a^{-2/7}\,(1-a^2)^{1/14}}\,\frac{\bbf (\brho)}{\brho^{1/4}}.
\label{exp:psi2InOut}
\ee

\subsection*{Inner layer $E^{1/3}$} 

In the inner layer, $\chi_2$ is constant and equal to $\alpha^2 a^2/(4\,\om^2)$. 
The function $\psi _2$ has by contrast a more complex structure.    
As shown by \cite{stewartson1966}, $\psi _2= O(E^{19/42})$ in the inner layer. 
More precisely, it can be written as
\be
\psi _2 \sim E^{19/42} \,\Lambda_3(a,\om)\, \bI (R,Z) ,
\ee   
with 
\bsea 
R &=  &\frac{(\rho -a)}{(1-a^2)^{1/6}\,E^{1/3} }, \\
Z &= &\frac{z}{\sqrt{1-a^2 }} ,
\esea
and $\bI(R,Z)$ which satisfies 
\be
\frac{\partial ^6 \bI}{\partial R^6} +4 \,\pdd{\bI}{Z} =0 ,
\label{equ:I}
\ee
with 
\bsea  
(-2R)^{1/4}\,\bI  \sim 1 ~{\rm when} ~Z\rightarrow 0~{\rm for}~ R<0 , \\
(-2 R)^{1/4}\,\bI  \sim  (1-Z) ~{\rm when} ~  R\rightarrow -\infty ,\\
\bI (R,Z=1) =0 ,\\
\bI (R,Z=0) =0 ~{\rm for} ~ R>0  ,\\
\bI  \rightarrow 0 ~{\rm as}~{R\rightarrow +\infty} .
\esea 
 An integral form of  $\bI(R,Z)$ can easily be obtained from (\ref{equ:I}) using Fourier transform techniques as
\be
\bI (R,Z) =\frac{\Gamma(3/4) }{2^{1/4} \,\pi} \int_{0}^{\infty}\frac{\sinh\left[k^3(1-Z)/2\right]}{\sinh(k^3/2)} \frac{\cos(kR+3\pi/8)}{k^{3/4}}\, \mathrm{d}k .
\label{exp:I}
\ee

The dependence of $\psi _2$ with respect to $\om$ is only in the function $\Lambda_3$ which is found to
be 
\be
\Lambda_3(a,\om)=  -\frac{a^{3/28}}{2^{6/7}\,\bbf'(0)\,(1-a^2)^{11/84}} \,\delta \Bchi (a,\om) ~.
\label{exp:A3}
\ee

The axial flow $u_{z_2}$ which can be deduced in the inner layer from  $\psi _2$ by 
\be
u_{z_2}= \frac{1}{a}\pa{\psi_2}{\rho}
\ee
is then given by
\be
u_{z_2} \sim E^{5/12}\, \Lambda_4(a,\om)\, \bJ (R,Z) ,
\ee
with
\be
\Lambda_4(a,\om) = \frac{ \Lambda_3(a,\om)}{a\,(1-a^2)^{1/6}}=  -\frac{ \delta \Bchi (a,\om)}{2^{6/7}\,\bbf'(0)\,a^{25/28}(1-a^2)^{25/84}}, 
\label{exp:A4}
\ee
and 
\be
\bJ(R,Z) = \frac{\Gamma(3/4) }{2^{1/4} \,\pi} \int_{0}^{\infty}\frac{\sinh\left[k^3(1-Z)/2\right]}{\sinh(k^3/2)} \cos(\pi/8- k R)\,k^{1/4}\, \mathrm{d}k .
\label{exp:J}
\ee

~\\ 

For the comparison with the numerical results, both leading-order and second-order approximations are used. 
Second order approximations   can be obtained by considering the $O(E^{1/28})$ correction to $\Bchi ^-$ deduced from
(\ref{exp:chi2}): 
\be
\Bchi ^{-} = \frac{\alpha^2 a^2}{4\, \om^2} + E^{1/28} \Lambda_2\, \bbf (0). 
\ee
 This provides a new expression for $\delta \Bchi$:
 \be
 \delta \Bchi = \frac{\BF(a;\om) - \alpha^2 a^2/(4 \,\om^2) }{1+ \zeta\, E^{1/28}} ~,
\label{exp:Newchi}
\ee
with 
\be
\zeta = -\left(\frac{2}{a}\right)^{1/7}\frac{(1-a^2)^{-5/56}\,\bbf(0)}{\mathbf{f}'(0)}. 
\ee
If we use (\ref{exp:Newchi}) in the expressions (\ref{exp:A1}), (\ref{exp:A2}), (\ref{exp:A3}) and (\ref{exp:A4}) for the coefficients $\Lambda_1$, $\Lambda_2$, $\Lambda_3$, $\Lambda_4$, we obtain approximations 
valid up to $O(E^{1/14})$ corrections. It is these second-order approximations which have been used in figures \ref{fig:comparuphi}(b),  \ref{fig:comparwz}(b) and \ref{fig:comparuz}(b).

\end{document}